\documentclass[10pt]{article}
\usepackage[margin=.8in,left=.8in]{geometry}

\usepackage{amsthm}
\usepackage{amssymb}
\usepackage{amsmath}
\usepackage{mathtools}
\usepackage{adjustbox}

\usepackage[table]{xcolor}

\usepackage{stmaryrd}    
\usepackage{rotating}
\usepackage{xfrac}
\usepackage{enumitem}\setlist{nosep}
\usepackage{slashed}

\usepackage[safe]{tipa} 

\usepackage{pgffor}

\usepackage{hhline}

\usepackage{tensor}

\usepackage{accents}
\newlength{\dhatheight}

\usepackage{tikz}
\usetikzlibrary{
  backgrounds, 
  decorations.pathmorphing, 
  decorations.markings,
  decorations.text,
  snakes
}
\usetikzlibrary{cd}

\usepackage{mathrsfs} 
\DeclareMathAlphabet{\mathpzc}{OT1}{pzc}{m}{it} 

\usepackage{hyperref}
\hypersetup{
  colorlinks = true,
  linkcolor=darkblue, citecolor=darkblue,
  urlcolor=black
}

\definecolor{darkblue}{rgb}{0.05,0.25,0.65}
\definecolor{darkgreen}{RGB}{20,140,10}
\definecolor{lightgray}{rgb}{0.9,0.9,0.9}
\definecolor{darkorange}{RGB}{200,100,5}
\definecolor{darkyellow}{rgb}{.91,.91,0}

\theoremstyle{definition}

\newcommand{\acts}{\raisebox{1.4pt}{\;\rotatebox[origin=c]{90}{$\curvearrowright$}}\hspace{.5pt}}

\newcommand{\rchi}{\raisebox{1pt}{$\chi$}}

\newcommand{\ZTwo}{\mathbb{Z}_2}

\let\PLAINthebibliography\thebibliography
\renewcommand\thebibliography[1]{
  \PLAINthebibliography{#1}
  \setlength{\parskip}{0.5pt}
  \setlength{\itemsep}{0.5pt plus .3ex}
}

\newcommand{\defneq}{\equiv}

\newcommand\bosonic[1]{\mathstrut\mkern2.5mu#1\mkern-14mu\raise1.7ex%
  \hbox{$\scriptstyle\rightsquigarrow$}}

\newcommand{\grayunderbrace}[2]{\mathcolor{gray}{\underbrace{\mathcolor{black}{#1}}}_{\mathcolor{gray}{#2}}}

\newcommand{\plus}{{\sqcup \{\infty\}}}
\newcommand{\cpt}{\hspace{.8pt}{\adjustbox{scale={.5}{.77}}{$\cup$} \{\infty\}}}

\begin{document}

\setlength{\abovedisplayskip}{3pt}
\setlength{\belowdisplayskip}{3pt}
\setlength{\abovedisplayshortskip}{-3pt}
\setlength{\belowdisplayshortskip}{3pt}

\title{Anyons on M5-Probes of Seifert 3-Orbifolds 
\\ 
via Flux Quantization}

\author{
  Hisham Sati${}^{\,\hyperlink{DoS}{a}, \hyperlink{Courant}{b}}$,
  \;\;
  Urs Schreiber${}^{\,\hyperlink{DoS}{a}}$
}

\maketitle

\begin{abstract}
  We observe that there is a rigorous derivation of (abelian) anyonic quantum states, hence of ``topological order'', on the 1+2-dimensional fixed locus of M5-probes wrapped over a trivially Seifert-fibered 3-orbifold singularity. 

\smallskip  
  Similar statements have previously been conjectured by appeal to the unknown dynamics of ``coincident'' M5-branes, but neglecting effects of flux-quantization that, as we highlight, entail anyonic solitons already in the rigorously tractable case of single M5-brane probes.

\smallskip 
   This is possible after globally completing the ``self-dual'' tensor field on probe M5-branes by flux-quantization in the non-abelian cohomology theory called {\it equivariant twistorial Cohomotopy}, which is admissible by recent results.

\end{abstract}

\vspace{.8cm}

\begin{center}
\begin{minipage}{8.5cm}
  \tableofcontents
\end{minipage}
\end{center}

\medskip

\vfill

\hrule
\vspace{5pt}

{
\hypertarget{DoS}{}
\footnotesize
\noindent
\def\arraystretch{1}
\tabcolsep=0pt
\begin{tabular}{ll}
${}^a$\,
&
Mathematics, Division of Science; and
\\
&
Center for Quantum and Topological Systems,
\\
&
NYUAD Research Institute,
\\
&
New York University Abu Dhabi, UAE.  
\end{tabular}
\hfill
\adjustbox{raise=-15pt}{
\href{https://ncatlab.org/nlab/show/Center+for+Quantum+and+Topological+Systems}{\includegraphics[width=3cm]{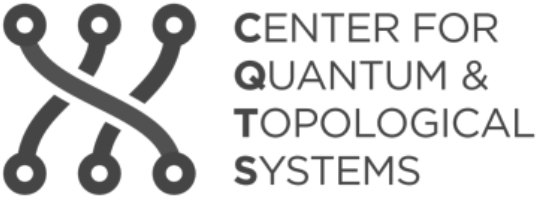}}
}

\vspace{1mm} 
\hypertarget{Courant}{}
\noindent 
\def\arraystretch{1}
\tabcolsep=0pt
\begin{tabular}{ll}
${}^b$\,The Courant Institute for Mathematical Sciences, NYU, NY.
\end{tabular}

\vspace{.2cm}

\noindent
The authors acknowledge the support by {\it Tamkeen} under the 
{\it NYU Abu Dhabi Research Institute grant} {\tt CG008}.
}

\newpage

\section{Introduction \& Overview}

Anyonic topological order (cf. \cite{ZCZW19}\cite{SS23-ToplOrder}) is arguably the holy grail of quantum materials research, potentially providing the much sought-after hardware for fault-tolerant scalable quantum computation via topological qbits (cf. \cite{Sau17}\cite{MySS24}). However, its microscopic understanding --- like that of all strongly-coupled/correlated quantum systems --- has remained sketchy at best, which may partly explain the current dearth of experimental realizations.

\smallskip

One candidate approach to a general understanding of strongly-coupled quantum systems is to map their dynamics to the fluctuations of higher-dimensional ``Membranes'' (whence: ``M-theory'' \cite{Duff96}\cite{Duff99-MTheory})  in an auxiliary higher-dimensional gravitational system (``geometric engineering'' or ``holographic duality'', cf. \cite{HKSS07}\cite{ZLSS15}\cite{HLS18}). Specifically 5-dimensional such branes (``M5-branes'', cf. \cite[\S 3]{Duff99-MTheory}\cite{GSS24-FluxOnM5}\cite{GSS24-M5Embedding}) have more recently been argued 
\cite{ChoGangKim20} \footnote{
  The relation conjectured in \cite{ChoGangKim20} between Seifert 3-folds and modular tensor categories (topological order)  has been further discussed by \cite{CGQZ22}\cite{CQW23}\cite{BSNW24}, but in the abstract, without further attention to the conjectured origin in M5-brane physics.
}
to exhibit topological order when their worldvolume is fibered in (``wrapped on'') Seifert 3-manifolds \cite{LeeRaymond01}, as considered in \cite{GukovPei17}, building on a conjecture known as the ``3d-3d correspondence''  (cf. \cite{Dimofte15}).

\smallskip

Unfortunately,  an actual formulation of M-theory, beyond a tantalizing web of hints and plausibility checks, remains an open problem (cf. \cite[p. 43]{Moore14}\cite{Duff20}).
Hence, these arguments have also largely remained informal (conjectural), not the least because they refer to the would-be non-abelian worldvolume theory of $N > 1$ ``coincident'' M5-branes, which remains notoriously undefined (cf. \cite{FreedTeleman14}), particularly so in the relevant sector of non-perturbative field strength where the worldvolume tensor field ceases to be self-dual (cf. \cite[Rem. 3.19]{GSS24-FluxOnM5}).

\smallskip

But to proceed to a more precise formulation of actual M-theoretic effects, we have previously highlighted that: 

\vspace{0.5mm}
\begin{itemize}

\item[\bf (i)] the first non-perturbative phenomenon to take care of is the global topological completion of the usual local field content on M-branes by {\it flux-quantization laws} \cite{SS24-Flux};

\item[\bf (ii)] the admissible such laws in M-theory are necessarily {\it non-abelian} generalized cohomology theories \cite{FSS23-Char} -- due to the
{\it non-linear} Bianchi identity
\eqref{SuperBianchiAndEOMOf11DSuGra}
of the C-field in 11D SuGra, cf. \cite[\S 2.5]{Sati13}\cite{SS23-PhaseSpace}\cite{GSS24-SuGra}\cite[\S 4.2]{SS24-Flux} -- which have found little attention before;

\item[\bf (iii)]
among the admissible such choices for M-theory is (unstable, twisted) {\it Cohomotopy} \cite{FSS20-H}\cite{GradySati21}\cite{FSS21-Hopf}\cite{SS23-MF} (which is {\it linearly approximated} by Topological Modular Forms \cite[Rem. 7.4]{FSS23-Char}) 
and its ``twistorial'' variant \cite{FSS22-Twistorial}\cite{SS25-EquTwistorial};

\item[\bf (iv)] implementing this non-abelian flux quantization reveals solitonic quantum states not otherwise visible,  among which generically are skyrmions and anyons on (single, $N=1$) M5-brane probes \cite{GSS24-FluxOnM5}.
\end{itemize}

\smallskip

Here we apply this argument specifically to M5-branes probing (Seifert-fibered) {\it orbifolds}, which means \cite{SS20-Orb} that the flux-quantizing cohomology theory is to be promoted to an {\it equivariant} cohomology theory (cf. \cite{SS21-EBund}). An admissible such equivariant candidate has been constructed before, called {\it equivariant twistorial Cohomotopy} \cite{SS25-EquTwistorial}. 

\smallskip

We explain here how this result combines with our previous derivation \cite{SS24-AbAnyons}\cite{SS25-TQBits} to imply
(abelian\footnote{
  There is an analogous though more involved construction \cite{SS23-DefectBranes} of {\it non-abelian} anyons in a scenario akin to {\it intersecting} M5-branes, but its rooting in 11D super-gravity is less clear at the moment. Here it may be noteworthy that abelian anyons are the only ones currently observed in experiment with some certainty \cite{NakamuraEtAl20},
  and that their potential role in topological quantum computation is much stronger \cite{Pachos06}\cite{Lloyd02}\cite{Wootton10}\cite{WoottonPachos11} than often appreciated.
}) anyonic quantum states and hence (abelian) ``topological order'' also on the 1+2-dimensional orbi-fixed locus of single M5-branes probing trivially-fibered Seifert $\ZTwo$-orbi 3-folds
(cf. \cite[Def. 2.6]{MecchiaSeppi20}, 
in fact super $\widehat{Z}_2$-orbifolds, see \S\ref{TwistorialFluxQuantization}):
\vspace{-1mm} 
\begin{equation}
  \label{TheBraneDiagram}
  \adjustbox{}{
  \begin{tikzcd}[
    column sep=-5pt,
    row sep=-20pt
  ]
  \scalebox{.7}{
    \color{darkblue}
    \bf
    \def\arraystretch{.85}
    \begin{tabular}{c}
      M5-brane probe
      \\
      worldvolume
    \end{tabular}
  }
  &&
  \scalebox{.7}{
    \color{darkblue}
    \bf
    \def\arraystretch{.85}
    \begin{tabular}{c}
      2-brane worldvolume
      \\
      hosting anyonic solitons
    \end{tabular}
  }
  &&
  \scalebox{.7}{
    \color{darkblue}
    \bf
    \def\arraystretch{.85}
    \begin{tabular}{c}
      M-theory
      \\
      circle
    \end{tabular}
  }
  &&
  \scalebox{.7}{
    \color{darkblue}
    \bf
    \def\arraystretch{.85}
    \begin{tabular}{c}
      cone
      \\
      orbifold
    \end{tabular}
  }
  \\[16pt]
  \Sigma^{1,6}
  &=&
  \mathbb{R}^{1,0}
  \times
  \mathbb{R}^2_{\cpt}
  &\times&
  S^1
  &\times&
  \mathbb{R}^2 \sslash \ZTwo
  \\[-12pt]
  &&
  \adjustbox{
    raise=3.8cm,
    scale=.7
  }{
    \begin{tikzpicture}
\begin{scope}[
  scale=.8,
  shift={(.7,-4.9)}
]

  \shade[right color=lightgray, left color=white]
    (3,-3)
      --
    (-1,-1)
      --
        (-1.21,1)
      --
    (2.3,3);

  \draw[dashed]
    (3,-3)
      --
    (-1,-1)
      --
    (-1.21,1)
      --
    (2.3,3)
      --
    (3,-3);

  \node[
    scale=1
  ] at (3.2,-2.1)
  {$\infty$};

  \begin{scope}[rotate=(+8)]
  \draw[dashed]
    (1.5,-1)
    ellipse
    (.2 and .37);
  \draw
   (1.5,-1)
   to 
    node[above, yshift=-1pt]{
     \;\;\;\;\;\;\;\;\;\;\;\;\;
     \rotatebox[origin=c]{7}{
     \scalebox{.7}{
     \color{darkorange}
     \bf
       anyon
     }
     }
   }
    node[below, yshift=+6.3pt]{
     \;\;\;\;\;\;\;\;\;\;\;\;\;\;\;
     \rotatebox[origin=c]{7}{
     \scalebox{.7}{
     \color{darkorange}
     \bf
       worldline
     }
     }
   }
   (-2.2,-1);
  \draw
   (1.5+1.2,-1)
   to
   (4,-1);
  \end{scope}

  \begin{scope}[shift={(-.2,1.4)}, scale=(.96)]
  \begin{scope}[rotate=(+8)]
  \draw[dashed]
    (1.5,-1)
    ellipse
    (.2 and .37);
  \draw
   (1.5,-1)
   to
   (-2.3,-1);
  \draw
   (1.5+1.35,-1)
   to
   (4.1,-1);
  \end{scope}
  \end{scope}
  \begin{scope}[shift={(-1,.5)}, scale=(.7)]
  \begin{scope}[rotate=(+8)]
  \draw[dashed]
    (1.5,-1)
    ellipse
    (.2 and .32);
  \draw
   (1.5,-1)
   to
   (-1.8,-1);
  \end{scope}
  \end{scope}

\end{scope}
    \end{tikzpicture}
    }
    &\times&
    \adjustbox{
      raise=3pt
    }{
    \begin{tikzpicture}
      \draw[
        line width=2.2,
        draw=darkgreen
      ]
        (0,0) circle (1);
    \end{tikzpicture}
    }
    &\times&
    \adjustbox{
      raise=3pt
    }{
\begin{tikzpicture}
\begin{scope}[
  xscale=.7,
  yscale=.5*.7
]
 \shadedraw[draw opacity=0, top color=darkblue, bottom color=cyan]
   (0,0) -- (3,3) .. controls (2,2) and (2,-2) ..  (3,-3) -- (0,0);
 \draw[draw opacity=0, top color=white, bottom color=darkblue]
   (3,3)
     .. controls (2,2) and (2,-2) ..  (3,-3)
     .. controls (4,-3.9) and (4,+3.9) ..  (3,3);
\end{scope}
\end{tikzpicture}
    }
  \end{tikzcd}
  }
\end{equation}
The point is that once the flux quantization law is admissibly chosen in equivariant twistorial Cohomotopy, we obtain the anyon quantum states  as a rigorous result instead of an informal conjecture --- a rare feat in M-theory.

\smallskip 
Since the heavy lifting of proving the main mathematical theorems that we use has been done in \cite{SS24-AbAnyons}\cite{SS25-EquTwistorial}, we focus here on explaining the physics background in super-space super-gravity and the conceptual steps in the construction of the flux quantization.

\section{Solitons via Flux Quantization}
\label{SolitonsViaFluxQuantization}

Before turning to the case of interest,  we briefly set the scene along familiar examples of solitons in Yang-Mills theory.

\medskip

\noindent
{\bf Algebraic Topology for Solitonic fields.} While the math we use in this article is classical --- basic algebraic topology and homotopy theory dating back over a century (cf. \cite{Hatcher02}\cite{Strom11}\cite{FomenkoFuchs16}\cite[\S 1]{FSS23-Char})  --- it may still not be as familiar to theoretical physicists. This is, in part, our point: contemporary gauge-theoretic analysis, even where developed to sophistication such as in BV-BRST formalism \cite{HenneauxTeitelboim92}, tends to be all concerned with {\it local} behavior of fields, their {\it infinitesimal analysis} such as expressed in differential equations of motion on differential forms such as flux densities. 

On the other hand, algebraic topology, historically conceived as the complementary {\it spatial analysis} (Poincar{\'e}'s {\it analysis situs} \cite{Poincare1895}), is the tool to describe the global behavior of solitonic fields, such as skyrmions or anyons. 
\footnote{
Both perspectives -- infinitesimal and spatial analysis -- unify in modern geometric homotopy theory of higher stacks \cite{FSS14-Stacky}\cite[\S 9]{FSS23-Char}\cite[\S 3.3]{SS24-Flux}, which however we try to avoid further mentioning here, for brevity.}

\smallskip

The basic concept of homotopy theory (and also of cohomology theory, as we will see in a moment) is that of {\it homotopy} (cf. \cite[\S 3.1]{FomenkoFuchs16}\cite[Fig. 4.2.2]{Marsh18}): Given a pair of topological spaces $X$ and $Y$, and given a pair of continuous maps $f_0, f_1 \,:\, X \xrightarrow{\;} Y$, a homotopy $\eta$ between them, suggestively denoted as
$$
    \begin{tikzcd}[
    ampersand replacement=\&,
    column sep=20pt
  ]
    X
    \ar[
      rr,
      bend left=27,
      "{ f_0 }"{description}
    ]
    \ar[
      rr,
      bend right=27,
      "{ f_1 }"{description}
    ]
  \&
    \mathclap{
    \hspace{1pt}
    \rotatebox[origin=c]{-90}{${\Rightarrow}$}{\;\scalebox{0.7}{$\eta$}}
    }
    \&
    Y
    \,,
  \end{tikzcd}
$$
is a continuous deformation of $f_0$ into $f_1$, namely a continuous 1-parameter family of such maps interpolating between them:
$$
  \eta
  \;:\;
  \begin{tikzcd}
  \mathbb{R}
  \times
  X 
  \ar[
    rr,
    "{
      \mathrm{cntns}
    }"
  ]
  &&
  Y
  \end{tikzcd}
  \,,
  \hspace{.6cm}
  \mbox{s.t.}
  \hspace{.6cm}
  \eta(0,x) \,=\, f_0(x)
  \,,
  \;\;\;
  \eta(1,x) \,=\, f_1(x)
  \,.
$$
Homotopy is an equivalence relation whose classes are the {\it homotopy classes} $[f]$ of maps (cf. \cite[\S 3.2]{FomenkoFuchs16}).

Said differently, in the {\it mapping space}
$\mathrm{Maps}(X,Y)$ of all continuous maps from $X$ to $Y$ (cf. \cite[Rem. 1.0.17]{SS21-EBund}), homotopies are the continuous paths, and homotopy classes $[f]$ are the path-connected components $\pi_0(-)$:
\begin{equation}
  \label{HomotopyClass}
  [f]
  \;\;
  \in
  \;\;
  \pi_0
  \, 
  \mathrm{Maps}(
    X,\, Y
  )
  \,.
\end{equation}
For instance, one says that the {\it $n$-th homotopy group} (cf. \cite[\S 4]{Nakahara03}\cite[\S 8]{FomenkoFuchs16}) of a connected space $X$ is, for $n \geq 1$, the homotopy classes \eqref{HomotopyClass} of maps from the $n$-sphere into it:
\begin{equation}
  \label{HomotopyGroups}
  \pi_n\big(   
    X
  \big)
  \;\;
  \defneq
  \;\;
  \pi_0
  \,
  \mathrm{Maps}\big(
    S^n
    ,\,
    X
  \big)
  \,,
\end{equation}
and one says that $f : X \xrightarrow{\;} Y$ is a {\it homotopy equivalence} (cf. \cite[\S 3.3]{FomenkoFuchs16}), so that $X \,\simeq\, Y$, if there exists reverse map $\overline{f} : Y \xrightarrow{} X$ which is an inverse up to homotopy, in that $[\overline{f} \circ f] \,=\, [\mathrm{id}_X]$ and $[f \circ \overline{f}] \,=\, [\mathrm{id}_Y]$.

\smallskip

A familiar example in quantum field theory is the classification of {\bf Yang-Mills instantons} in 4D (cf. \cite[\S 1.1]{Tong05}, equivalently the {\it solitons} of 5D YM, cf. \cite[\S 4]{PapageorgakisRoyston14}): The topological class of an $\mathrm{SU}(2)$-gauge field on $\mathbb{R}^4$ with field strength ``vanishing at infinity'', namely on a large enough 3-sphere $S^3 \subset \mathbb{R}^4$, is the homotopy class of the gauge transformation $g \,:S^3 \xrightarrow{\;} \mathrm{SU}(2)$ which trivializes the asymptotic gauge potential:
\begin{equation}
  \label{BPSTInstantonClassification}
  [g]
  \,\in\,
  \pi_0\,
  \mathrm{Maps}\big(
    S^3
    ,\,
    \grayunderbrace{
    \mathrm{SU}(2)
    }{
      \mathclap{
        \,\simeq\, S^3
      }
    }
  \big)
  \;\;
  \simeq
  \;\;
  \pi_0\,
  \mathrm{Maps}\big(
    S^3
    ,\,
    S^3
  \big)
  \;\;
  \defneq
  \;\;
  \pi_3(S^3)
  \;\;
  \simeq
  \;\;
  \mathbb{Z}
  \,.
\end{equation}
We will be considering just somewhat more sophisticated variants of this standard example, but first to recall the following more flexible perspective:

\medskip

\noindent
{\bf Solitonic fields vanishing at infinity.} Generally, the condition that and how  solitonic fields ``vanish at infinity'' is mathematically reflected by regarding all spaces $X$ as equipped with a chosen point $\infty \in X$, and requiring all maps between to respect this point, which we shall do in all of the following.

For instance, if we write $\mathbb{R}^n_{\cpt} \,\simeq\, S^n$ for the result of adjoining a single point at infinity to Euclidean space (technically: ``Alexandroff one-point compactification'', cf. \cite[(1) (17)]{SS23-MF}) then the space of pointed maps from this to any pointed space is the $n$-fold {\it based loop space}:
$$
  \Omega^n_\infty X
  \;\;
  \simeq
  \;\;
  \mathrm{Maps}\big(
    \mathbb{R}^{n}_{\cpt}
    ,\,
    (X,\infty)
  \big)
  \,.
$$
On the other hand, the $n$-sphere with a {\it disjoint} (unreachable) point-at-infinity is denoted $S^n_{\plus}$, with
$$
  \mathcal{L} X
  \;\;
    :=
  \;\;
  \mathrm{Maps}\big(
    S^1_\plus
    ,\,
    X
  \big)
$$
being the {\it free loop space}.

As another example, given a pair $(X,\infty)$ and $(Y,\infty)$, one clearly wants to regard in the product space $X \times Y$ all those points as being ``at infinity'' which are so with respect to either factor. This is accomplished by their ``smash product'' space
\begin{equation}
  \label{SmashProduct}
  X \wedge Y
  \;:=\;
  \frac{X \times Y}{
    X \!\!\times\!\! \{\infty\}
    \,\cup\,
    \{\infty\} \!\!\times\!\! Y
  }
  \,.
\end{equation}
For instance (cf. \cite[\S A.2]{SS23-Obs})
\begin{equation}
  \label{SmashOfSpheres}
  \mathbb{R}^{n}_{\cpt}
  \,\wedge\,
  \mathbb{R}^{m}_{\cpt}
  \;\;
  \simeq
  \;\;
  \mathbb{R}^{n+m}_{\cpt}
  \,,
  \hspace{1cm}
  \mathrm{Maps}\big(
    \mathbb{R}^n_{\cpt}
    \,\wedge\, X
    ,\,
    Y
  \big)
  \;\;
  \simeq
  \;\;
  \mathrm{Maps}\big(
    X
    ,\,
    \Omega^n_\infty Y
  \big)
  \,.
\end{equation}

\smallskip

This way, the above classification of Yang-Mills instantons/solitons 
\eqref{BPSTInstantonClassification}
becomes the more systematic statement that topological sectors of $\mathrm{SU}(2)$-gauge fields --- whose {\it classifying space} $B \mathrm{SU}(2)$ is a connected space with $\Omega B \mathrm{SU}(2) \,\simeq\, \mathrm{SU}(2)$, cf. \cite{RudolphSchmidt17} ---  on $\mathbb{R}^4$ are derived alternatively by:
\vspace{-2mm} 
$$
  \def\arraystretch{1.7}
  \def\arraycolsep{2pt}
  \begin{array}{l}
  \pi_0
  \,
  \mathrm{Maps}\big(
    \mathbb{R}^4_{\cpt}
    ,\,
    B \mathrm{SU}(2)
  \big)
  \;\simeq\;
  \pi_0
  \,
  \mathrm{Maps}\big(
    \mathbb{R}^3_{\cpt} \wedge \mathbb{R}^1_{\cpt}
    ,\,
    B \mathrm{SU}(2)
  \big)
  \;\simeq\;
  \pi_0
  \,
  \mathrm{Maps}\big(
    S^3
    ,\,
    \Omega
      B \mathrm{SU}(2)
  \big)
  \\
\hspace{4.06cm}   \;\simeq\;
  \pi_0
  \,
  \mathrm{Maps}\big(
    S^3
    ,\,
    \mathrm{SU}(2)
  \big)
  \;\simeq\;
  \mathbb{Z}
  \,.
  \end{array}
$$

\smallskip

\noindent
{\bf Electromagnetic flux quantization.}
Essentially by replacing the group $\mathrm{SU}(2)$ in \eqref{BPSTInstantonClassification} by $\mathrm{U}(1)$,
it is a classical and famous fact that the ordinary electromagnetic field is to be ``flux-quantized'' in integral 2-cohomology (``Dirac charge quantization'', cf. \cite{Alvarez85}). First of all this means that the electromagnetic flux density, a differential 2-form $F_2$ on spacetime $X$ (cf. \cite[\S3.5 \& 7.2b]{Frankel97}\cite[\S 3.1]{SS23-PhaseSpace}), is to be accompanied by an integral cohomology class ${[\rchi]}$ that lifts its de Rham class through the de Rham homomorphism
(cf. \cite[Thm. 8.9 \& Prop. 10.6]{BottTu82})
from integral cohomology (which we call the {\it character map} on ordinary cohomology \cite[Ex. 7.1]{FSS23-Char}):
\begin{equation}
  \label{OrdinaryCharacterMap}
  \begin{tikzcd}[
    row sep=0pt
  ]
    H^2(X;\, \mathbb{Z})
    \ar[
      rr,
      "{
        (\mathbb{Z}
        \hookrightarrow 
        \mathbb{R})_\ast
      }"
    ]
    \ar[
      rrrr,
      rounded corners,
      to path={
             ([yshift=+00pt]\tikztostart.north)    
          -- ([yshift=+10pt]\tikztostart.north)    
          -- node[yshift=5pt]{
            \scalebox{.7}{
              \color{darkgreen}
              \bf
              character map on integral cohomology
            }
          }
            node[yshift=-5pt]{
              \scalebox{.7}{$
              \mathrm{ch}
            $}}
             ([yshift=+10pt]\tikztotarget.north)    
          -- ([yshift=+00pt]\tikztotarget.north)    
      }
    ]
    &&
    H^2(X;\, \mathbb{R})
    \ar[
      rr,
      "{ \sim }"
    ]
    &&
    H^2_{\mathrm{dR}}(X)
    \\
    \mathllap{
      \scalebox{.7}{
        \color{darkblue}
        \bf
        \def\arraystretch{.9}
        \begin{tabular}{c}
          soliton
          \\
          number
        \end{tabular}
      }
    }
    {[\rchi]}
    \ar[
      rrrr,
      |->,
      shorten=25pt
    ]
    &&&&
    {[F_2]}
    \mathrlap{
      \scalebox{.7}{
        \color{darkblue}
        \bf
        \def\arraystretch{.9}
        \begin{tabular}{c}
          class of 
          \\
          flux density
        \end{tabular}
      }
    }
  \end{tikzcd}
\end{equation}
(From this perspective, the system of local gauge potentials $\widehat{A}$ of the electromagnetic field constitutes a coboundary $\rchi \Rightarrow F_2$ in real cohomology \cite[Prop. 9.5]{FSS23-Char}, but here we need not further dwell on this aspect.)

\smallskip

This implies that the ``branes'' of electromagnetism, namely the magnetic monopoles (hypothetically) as well as the Abrikosov vortices in type II superconductors (experimentally observed) come in integer units (cf. \cite[\S 2.1]{SS24-Flux}): For the monopole with worldline $\mathbb{R}^{1,0} \xhookrightarrow{\;} \mathbb{R}^{1,3}$ (the {\it singular brane} of electromagnetism),
this follows from the integral cohomology group of the spacetime {\it complement} of the singular locus
$$
  H^2\big(
    \grayunderbrace{
    \mathbb{R}^{1,3}
    \setminus
    \mathbb{R}^{1,0}
    }{
      \mathclap{
      \mathbb{R}^{1,0}
      \,\times\, 
      \mathbb{R}^1_{> 0}
      \,\times\,
      S^2
      \mathrlap{
      \;\,
        \underset{
          \mathclap{
            \mathrm{hmtp}
          }
        }{\simeq}
      \;\,
      S^2
      }
      }
    }
    ;\,
    \mathbb{Z}
  \big)
  \;\;
  \simeq
  \;\;
  H^2(
    S^2
    ;\,
    \mathbb{Z}
  )
  \;\;
  \simeq
  \;\;
  \mathbb{Z}
  \,,
$$
while for the Abrikosov vortices (the {\it solitonic branes} of electromagnetism) with worldsheets $\mathbb{R}^{1,1} \xhookrightarrow{} \mathbb{R}^{1,3}$ 
this follows from the integral cohomology group of the result of adjoining the {\it point-at-infinity} to their transverse space:
$$
  H^2\big(
    \grayunderbrace{
    \mathbb{R}^{1,1}
    \times
    \mathbb{R}^{2}_{\cpt}
    }{
      \mathclap{
        \underset{
          \mathclap{\mathrm{hmpt}}
        }{\simeq}
        \;\,
        S^2
      }
    }
    ;
    \,
    \mathbb{Z}
  \big)
  \;\;
  \simeq
  \;\;
  H^2(
    S^2
    ;\,
    \mathbb{Z}
  )
  \;\;
  \simeq
  \;\;
  \mathbb{Z}
  \,.
$$

\noindent
{\bf Classifying space for electromagnetic charges.}
Even more classical, but maybe less widely appreciated, is the fact (cf. \cite[Ex. 2.1]{FSS23-Char}) that also integral 2-cohomology has a {\it classifying space} $B \mathrm{U}(1) \,\simeq\,\mathbb{C}P^\infty$ (the union of all the finite-dimensional complex-projective spaces $\mathbb{C}P^n$   \cite[Ex. 0.6]{Hatcher02}, also called the ``Eilenberg-MacLane space'' $K(\mathbb{Z},2)$ \cite[p 139]{FomenkoFuchs16}), which is such that the 2-cohomology group of $X$ is naturally equivalent to the homotopy classes \eqref{HomotopyClass}
of continuous maps from $X$ into this classifying space (cf. \cite[p 263]{FomenkoFuchs16}\cite[Ex. 2.1]{FSS23-Char}):
\begin{equation}
  \label{IntegralCohohomologyRepresented}
  H^2(
    X
    ;\,
    \mathbb{Z}
  )
  \;\;
  \simeq
  \;\;
  \pi_0 
  \,
  \mathrm{Maps}(
    X
    ,\,
    B\mathrm{U}(1)
  )
  \,.
\end{equation}
We may read this as saying that a topological-sector field configuration $\Phi$ of the EM-field is a map $X \xrightarrow{\;} B\mathrm{U}(1)$ and that a gauge transformation $\Phi \Rightarrow \Phi'$ is a homotopy between such maps, so that the cohomology \eqref{IntegralCohohomologyRepresented} is the {\it gauge equivalence classes of topological field configurations}
\begin{equation}
  \label{ClassifyingEMSectors}
    \begin{tikzcd}[
    ampersand replacement=\&,
    column sep=15pt
  ]
    X
    \ar[
      rr,
      bend left=25,
      "{ \Phi }"{description}
    ]
    \ar[
      rr,
      bend right=25,
      "{ \Phi' }"{description}
    ]
  \&
    \mathclap{
    \hspace{12pt}
    \rotatebox[origin=c]{-90}{$\;\overset{\sim}{\Rightarrow}$}
    }
    \&
    B\mathrm{U}(1)
    \,.
  \end{tikzcd}
\end{equation}
But the classifying space carries more information than just about the topological sectors of fields, it also knows the topological classes of their large {\it gauge transformations}, which by \eqref{ClassifyingEMSectors} form the space of based loops $\Omega(-)$ in the mapping space, identified with the classes of $\mathrm{U}(1)$-valued maps:
\vspace{-1mm} 
\begin{equation}
  \label{GaugeClasses}
  \Omega_\Phi
  \,
  \mathrm{Maps}(
    X
    ,\,
    B\mathrm{U}(1)
  )
  \,
  \simeq
  \,
  \Omega_0
  \,
  \mathrm{Maps}(
    X
    ,\,
    B\mathrm{U}(1)
  )
  \,
  \simeq
  \,
  \mathrm{Maps}(
    X
    ,\,
    \Omega_0
    \,
    B\mathrm{U}(1)
  )
  \,
  \simeq
  \,
  \mathrm{Maps}\big(
    X
    ,\,
    \mathrm{U}(1)
  \big)
  \;
  \xrightarrow{ [-] }
  \;
  H^1(X;\, \mathbb{Z})
  \,.
\end{equation}

\newpage 
As such, the whole mapping space $\mathrm{Maps}(X,\, B\mathrm{U}(1))$ plays the role of the topological sector of the integrated on-shell {\it BRST-complex} of
electromagnetism (in fact, it is the shape of the {\it phase space stack} \footnote{
Just for readers with tolerance for stacks, we mention briefly that 
the picture \eqref{ClassifyingEMSectors}
of the topological sector of electromagnetism enhances to a full description of dynamical electromagnetism by \cite{SS24-Flux}
enhancing the space $B\mathrm{U}(1)$ to a {\it smooth stack} 
\cite{FSS14-Stacky}
$\mathbf{B} \mathrm{U}(1)_{\mathrm{conn}}$ (of which it is the underlying ``shape''), in which case maps of stacks are in bijection to consistent local 
electromagnetic gauge potentials $\widehat{A}$ and homotopies of maps of stacks are in bijection to $\mathrm{U}(1)$-gauge transformations between these,
$
  \begin{tikzcd}[
    ampersand replacement=\&,
    column sep=15pt
  ]
    X
    \ar[
      rr,
      bend left=25,
      "{ \widehat{A} }"{description},
    ]
    \ar[
      rr,
      bend right=20,
      "{ \widehat{A'} }"{description, pos=.55},
    ]
    \&
    \mathclap{
    \hspace{12pt}
    \rotatebox[origin=c]{-90}{$\;\overset{\sim}{\Rightarrow}$}
    }
    \&
   \mathbf{B}\mathrm{U}(1)_{\mathrm{conn}}
    \,,
  \end{tikzcd}
$
whose equivalence classes form the {\it differential} integral 2-cohomology
$
  H^2\big(
    X;\,\mathbb{Z}
  \big)_{\mathrm{diff}}
  \;\;
  \simeq
  \;\;
  \pi_0 
  \,
  \mathrm{Maps}
  \big(
    X
    ,\,
    \mathbf{B}\mathrm{U}(1)_{\mathrm{conn}}
  \big)
  \,.
$
While we do not further dwell on stacks and differential cohomology here, it is important that the discussion we provide concerns, in a precise sense, the topological sector of actual dynamical higher gauge fields, given by enhancing the non-abelian cohomology theories we discuss to differential cohomology.
} 
of electromagnetism, see \cite[\S 3.1]{SS23-PhaseSpace}). By the usual nature of observables in BRST theory (e.g. \cite[Def. 4.3]{Jiang23}), this means \cite{SS23-Obs}
that the {\it topological observables} on the flux-quantized electromagnetic field over a spacetime domain $X$ constitute the complex homology (cf. \cite[\S 2]{Hatcher02}\cite[\S 2]{FomenkoFuchs16}) 
of the mapping space into the classifying space
for the flux quantization:
\begin{equation}
  \label{ElectromagneticObservables}
  \mathrm{Obs}_\bullet
  \;:=\;
  H_\bullet\big(
    \mathrm{Maps}(
      X
      ,\,
      B\mathrm{U}(1)
    )
    ;\,
    \mathbb{C}
  \big)
  \,.
\end{equation}
The homological degree here is to be understood as the ``ghost''-degree of BRST theory: 
In degree=0 these topological observables are the electromagnetic soliton number
\eqref{OrdinaryCharacterMap}
$$
  \mathrm{Obs}_0
  \;\;
  \defneq
  \;\;
  H_0\big(
    \mathrm{Maps}(
      X
      ,\,
      B\mathrm{U}(1)
    )
    ;\,
    \mathbb{C}
  \big)
  \;\;
  =
  \;\;
  \mathbb{C}\big[
    \pi_0
    \,
    \mathrm{Maps}(
      X
      ,\,
      B\mathrm{U}(1)
    )
  \big]
  \;\;
  =
  \;\;
  \mathbb{C}\big[
    H^2(X;\mathbb{Z})
  \big]
  \,,
$$
and in degree=1 their classes 
\eqref{GaugeClasses}
of auto gauge-equivalences: 
\footnote{
  For $S$ a set, we write $\mathbb{C}[S]$ for its complex-linear span. If $S$ is equipped  
  with the structure of a group $G$, then this is the usual notation for the corresponding group-algebra, cf. \cite{Passman76}.
}
$$
  \mathrm{Obs}_1
  \;\;
  \defneq
  \;\;
  H_1\big(
    \mathrm{Maps}(
      X
      ,\,
      \mathbb{C}P^\infty
    )
    \;\,
    \mathbb{C}
  \big)
  \;\;
  \simeq
  \;\;
  \underset{
    \mathclap{H^2(X;\, \mathbb{Z})}
  }{\bigoplus}
  \;
  \mathbb{C}
  \big[
    H^1(X;\, \mathbb{Z})
  \big]
  \,.
$$

\smallskip

\noindent
{\bf RR-Field Flux Quantization.}
More conjectural but maybe also more famous than the above electromagnetic flux quantization in ordinary integral cohomology is the idea 
(\cite{MinasianMoore97}\cite{Witten98}\cite{BouwknegtMathai01}, cf. \cite{GS22}\cite[\S 4.1]{SS24-Flux})
to flux-quantize the RR-fields of type IIA supergravity in the ``extraordinary'' cohomology theory known as {\it twisted topological K-theory}. 

In hindsight, this is motivated by the observation that the Bianchi identities of the NS/RR-flux densities, 
$$
  \mathrm{d}
  \,
  H_3
  \;=\;
  0
  \,,
  \;\;\;\;\;\;
  \mathrm{d}
  \,
  F_{2\bullet}
  \;=\;
  H_3 \, F_{2\bullet-2}
  \,,
$$
are just the same kind of differential equations as satisfied in the image of the {\it twisted Chern character} map from twisted K-theory to
$H_3$-twisted de Rham cohomology (\cite{RohmWitten86}, cf. \cite[Ex. 6.6]{FSS23-Char}). This means, as in \eqref{OrdinaryCharacterMap}, that we may consider
completing the NS-field by a class $[\tau]$ in integral 3-cohomology and the  RR-field by a class $[\rchi]$ in $\tau$-twisted K-theory whose 
Chern character coincides with the class of the RR-flux densities:
$$
  \begin{tikzcd}[row sep=-2pt, 
    column sep=0pt
  ]
    H^3(X;\, \mathbb{Z})
    \ar[
      rr
    ]
    &&
    H^3_{\mathrm{dR}}(X)
    \\
    {[\tau]}
    &\longmapsto&
    {[H_3]}
  \end{tikzcd}
  \hspace{1cm}
  \begin{tikzcd}[
    row sep=-2pt,
    column sep=40pt
  ]
    K^{\tau}(X)
    \ar[
      rr,
      "{
        \scalebox{.7}{
        \color{darkgreen}
        \bf
        \def\arraystretch{.8}
        \begin{tabular}{c}
          twisted
          Chern character
        \end{tabular}
        }
      }"
    ]
    &&
    H^{H_3}_{\mathrm{dR}}(
      X
    )\;.
    \\
    {[\rchi]}
    &\longmapsto&
    {[F_{2\bullet}]_{H_3}}
  \end{tikzcd}
$$
Where ordinary electromagnetic flux quantization stabilizes Abrikosov vortices, such RR flux quantization in K-theory is understood to stabilize non-supersymmetric D-brane solitons.
Instructive for our purpose is the observation that also twisted K-theory has a classifying space, namely (cf. \cite[Ex. 1.3.19, 4.5.4]{SS21-EBund})
the 
space $\mathrm{Fred}$ of (odd, self-adjoint) Fredholm operators (on any $\ZTwo$-graded countably infinite-dimensional complex Hilbert space) 
homotopy-quotiented by the conjugation action of the corresponding projective unitary group
$$
  \underset{
    \mathclap{
      \tau \in H^3(X;\, \mathbb{Z})
    }
  }{\bigoplus}
  \;
  K^\tau(X)
  \;\;\;
  \simeq
  \;\;\;
  \pi_0\, 
  \mathrm{Maps}\big(
    X
    ,\,
    \mathrm{Fred}
    \!\sslash\!
    \mathrm{PU}
  \big)
  \,.
$$

\noindent
{\bf General Flux Quantization.}  
These examples make it clear how flux quantization works in general \cite{SS24-Flux}:
\begin{center}
\colorbox{lightgray}{\parbox{0.5\textwidth}{
\it
  Flux quantization globally completes higher gauge theories
  \\
  by determining the stable solitonic field configurations.
}}
\end{center}
\vspace{-3.2mm} 
\begin{center}
\colorbox{lightgray}{\parbox{0.67\textwidth}{\it
Admissible flux quantization laws are generalized cohomology theories whose 
\\
character map takes values in solutions of the given Bianchi identities.
}}
\end{center}
\vspace{-3.5mm} 
\begin{center}
\colorbox{lightgray}{\parbox{0.66\textwidth}{\it
Generalized cohomology is homotopy classes of 
maps into a classifying space.
}}
\end{center}
\begin{equation}
  \label{GeneralizedCharacterMap}
  \hspace{-2cm} 
  \begin{tikzcd}[
    row sep=-3pt, column sep=large
  ]
    \mathllap{
      \scalebox{.7}{
        \color{darkblue}
        \bf
        \def\arraystretch{.86}
        \begin{tabular}{c}
          Generalized cohomology
          \\
          classified by space $\mathcal{A}$
        \end{tabular}
      }
    }
    H^1\big(
      X;\,
      \Omega \mathcal{A}
    \big)
    \ar[
      rr,
      "{
        \scalebox{.7}{
          \color{darkgreen}
          \bf
          \def\arraystretch{.9}
          \begin{tabular}{c}
            generalized
            \\
            character map
          \end{tabular}
        }
      }",
      "{
        \mathrm{ch}
      }"{swap}
    ]
    &&
    H^1_{\mathrm{dR}}(X;\, \mathfrak{l}\mathcal{A})
    \mathrlap{
      \scalebox{.7}{
        \color{darkblue}
        \bf
        \def\arraystretch{.86}
        \begin{tabular}{c}
          nonabelian de Rham cohomology
          \\
          with coefficients in 
          WH $L_\infty$ algebra
        \end{tabular}
      }
    }
    \\
    \mathllap{
      \scalebox{.7}{
        \color{darkblue}
        \bf
        \def\arraystretch{.86}
        \begin{tabular}{c}
          topological charge of
          \\
          field configuration
        \end{tabular}
      }
    }
    {[\rchi]}
    \ar[
      rr,
      |->,
      shorten=10pt
    ]
    &&
    \big[
      \vec F
      \;\vert\;
      \mathrm{d}\, F^i \,=\,
      P^i(\vec F\,)
    \big]
    \mathrlap{
      \scalebox{.7}{
        \color{darkblue}
        \bf
        \def\arraystretch{.86}
        \begin{tabular}{c}
          class of flux densities
          \\
          satisfying given Bianchi
        \end{tabular}
      }
    }
  \end{tikzcd}
\end{equation}

\medskip

\noindent
{\bf Plan.}
While we see that flux quantization is the global completion of higher gauge-field theories, needed for defining and identifying their solitonic field configurations, it has previously received little attention beyond the above two examples and their minor variants. 

\smallskip 
Our aim now is to have a close look at the admissible flux quantization laws for the ``self-dual'' tensor field on M5-brane probes (cf. \cite[\S 4.3]{SS24-Flux}\cite{GSS24-FluxOnM5}) in the case that their worldvolume wraps a (Seifert) orbi-singularity, and to show that there is a natural choice which implies (predicts) solitonic configurations with anyonic quantum states.

\smallskip

\section{Flux Quantization on M5-branes}

With admissible flux quantization laws reflecting the given Bianchi identities via their character images
\eqref{GeneralizedCharacterMap},
it is worth re-considering where flux Bianchi identities ``come from''. Remarkably, the Bianchi identities of supergravity and of super $p$-brane probes are ingrained in the structure of super-space:

\medskip

\noindent
{\bf Bulk C-Field in 11D SuGra.}
Namely a field configuration of 11D super-gravity is a {\it super}torsion-free super-coframe field $(E, \Psi)$
on 11D super-spacetime, and its Einstein-Rarita-Schwinger equations of motion are {\it equivalent} 
\cite[Thm. 3.1]{GSS24-SuGra} (following \cite{CF80}\cite{BrinkHowe80}\cite[\S III.8.5]{CDF91})
simply to the statement that flux super-densities of the following form
\begin{equation}
  \label{11DSuperFluxDensities}
  \def\arraystretch{1.5}
  \begin{array}{ccccl}
  G^s_4 
  &\defneq&
  G_4 \,+\, G_4^0
  &\defneq&
  (G_4)_{a_1 \cdots a_4}
  E^{a_1}\cdots E^{a_4}
  \,+\,
  \tfrac{1}{2}\big(
    \overline{\Psi}
    \;\Gamma_{a_1 a_2}\,
    \Psi
  \big)
  E^{a_1} E^{a_2}
  \,,
  \\
  G^s_7 
  &\defneq&
  G_7 \,+\, G_7^0
  &\defneq&
  (G_7)_{a_1 \cdots a_7}
  E^{a_1}\cdots E^{a_7}
  \,+\,
  \tfrac{1}{5!}\big(
    \overline{\Psi}
    \;\Gamma_{a_1 \cdots a_5}\,
    \Psi
  \big)
  E^{a_1} \cdots E^{a_5}
  \end{array}
\end{equation}
satisfy the usual form of the Bianchi identity, but on super-space 
\footnote{
In particular, the equations of motion \eqref{SuperBianchiAndEOMOf11DSuGra} include the Hodge duality relation $G_7 \;=\; \star \, G_4$ over the underlying ordinary spacetime.
}
\begin{equation}
  \label{SuperBianchiAndEOMOf11DSuGra}
  \left.
  \adjustbox{raise=1pt}{
  \begin{tikzcd}[
    row sep=-1pt,
    column sep=0pt
  ]
  \mathrm{d}\, G_4^s
  &=&
  0
  \\
  \mathrm{d}\, G_7^s
  &=&
  \tfrac{1}{2}G^s_4\, G_4^s
  \end{tikzcd}
  }
  \!\!\! \right\}
  \hspace{.9cm}
  \Leftrightarrow
  \hspace{.9cm}
  \left\{\!\!\!\!\!\!
  \scalebox{.9}{
    \def\arraystretch{.9}
    \begin{tabular}{c}
      Equations of Motion
      \\
      of 11D Supergravity
    \end{tabular}
  }
  \right.
\end{equation}
One of many subtle reasons behind this theorem is the quartic Fierz identities on 11D commuting spinors (\cite[(3.28)]{DF82}, cf. \cite[Prop. 2.73]{GSS24-SuGra}), which say that the super-fluxes \eqref{11DSuperFluxDensities}
satisfy their Bianchi identities 
\eqref{SuperBianchiAndEOMOf11DSuGra}
for vanishing ordinary flux, hence when $(G^s_4,\, G^s_7) \,=\, (G_4^0,\, G_7^0)$
\begin{equation}
  \def\arraystretch{1.8}
  \begin{array}{l}
  \mathrm{d}
  \Big(
  \tfrac{1}{2}
  \big(\hspace{1pt}
    \overline{\Psi}
    \,\Gamma_{a_1 a_2}\,
    \Psi
  \big)
  E^{a_1} E^{a_2}
  \Big)
  \;=\;
  0
  \\
  \mathrm{d}
  \Big(
  \tfrac{1}{5!}
  \big(\hspace{1pt}
    \overline{\Psi}
    \,\Gamma_{a_1 \cdots a_5}\,
    \Psi
  \big)
  E^{a_1} \cdots E^{a_5}
  \Big)
  \;=\;
  \tfrac{1}{2}
  \Big(
  \tfrac{1}{2}
  \big(\hspace{1pt}
    \overline{\Psi}
    \,\Gamma_{a_1 a_2}\,
    \Psi
  \big)
  E^{a_1} E^{a_2}
  \Big)
  \Big(
  \tfrac{1}{2}
  \big(\hspace{1pt}
    \overline{\Psi}
    \,\Gamma_{a_1 a_2}\,
    \Psi
  \big)
  E^{a_1} E^{a_2}
  \Big)
  \mathrlap{\,.}
  \end{array}
\end{equation}

Now, the most fundamental 
\footnote{
  The precise sense in which 4-Cohomotopy \eqref{CohomotopyCharacter} is the {\it most fundamental} choice of flux quantization of \eqref{SuperBianchiAndEOMOf11DSuGra} is that its classifying space admits a CW-complex structure with the {\it smallest number of cells}: namely consisting only of (a single 0-cell and) a single 4-cell.
}
flux quantization law, which is admissible for these non-linear Bianchi identities 
\eqref{SuperBianchiAndEOMOf11DSuGra}
turns out to be 4-{\it Cohomotopy} (cf. \cite[(27)]{SS24-Flux}), whose classifying space is the 4-sphere:
\begin{equation}
  \label{CohomotopyCharacter}
  \begin{tikzcd}[
    row sep=-2pt, column sep=30pt
  ]
    \pi^4(X)
    \,:=\,
    \pi_0
    \,
    \mathrm{Maps}\big(
      X
      ,\,
      S^4
    \big)
    \ar[
      rr,
      "{
        \mathrm{ch}
      }"
    ]
    &&
    H^1_{\mathrm{dR}}\big(
      X
      ;\,
      \mathfrak{l}S^4
    \big)
    \\
    \mathllap{
      \scalebox{.7}{
        \color{darkblue}
        \bf
        \def\arraystretch{.9}
        \begin{tabular}{c}
          topological charge
          \\
          of C-field in bulk
        \end{tabular}
      }
    }
    {[
      c
    ]}
    &\longmapsto&
    {\footnotesize
    \left[
    \def\arraystretch{.9}
    \def\arraycolsep{0pt}
    \begin{array}{l}
      \mathrm{d}\, G_7 \,=\,
      \tfrac{1}{2} G_4\, G_4
      \\
      \mathrm{d}\, G_4 \,=\, 0
    \end{array}
    \right]
    }
  \end{tikzcd}
\end{equation}
The assumption (\cite[\S 2.5]{Sati13}, ``Hypothesis H'') that (tangentially twisted) 4-Cohomotopy is the ``correct'' flux quantization of 11D supergravity for purposes of M-theory 
provably implies a whole list of subtle topological effects expected in M-theory \cite{FSS20-H}\cite{FSS21-Hopf}\cite{GradySati21}\cite{SS21M5Anomaly}, among them the shifted half-integrality of the 4-flux \cite[Prop. 3.13]{FSS20-H}. Therefore, here we do assume this 4-Cohomotopical flux quantization of 11D SuGra.

\medskip

\noindent
{\bf Self-dual B-field on M5.}
A similar miracle occurs for the M5-brane when formulated in super-space: A $\sfrac{1}{2}$BPS immersion of an M5 super-worldvolume into 11D super-spacetime
(\cite[Def. 2.19]{GSS24-FluxOnM5}, essentially the ``super-embedding'' of \cite{HoweSezgin97}\cite[\S 5.2]{Sorokin00})
$$
  \begin{tikzcd}
    \Sigma^{1,5 \,\vert\, 2 \cdot \mathbf{8}_+}
    \ar[
      rr,
      "{ \phi }",
      "{
        \scalebox{.7}{
          \color{gray}
          $\sfrac{1}{2}$BPS
        }
      }"{swap}
    ]
    &&
    X^{1,10\,\vert\,\mathbf{32}}
  \end{tikzcd}
$$
entails a non-linearly self-dual
\cite[Rem. 3.19]{GSS24-FluxOnM5}
3-flux density super-form $H_3^s$ on $\Sigma$
$$
  H^s_3 
  \;\defneq\;
  (H_3)_{a_1 a_2 a_3}
  \,
  e^{a_1} \, e^{a_2}\, e^{a_3}
$$
whose Bianchi identity makes it a coboundary for the pullback of the 4-flux \cite[Prop. 3.18]{GSS24-FluxOnM5}, 
\begin{equation}
  \label{GaussLawOnM5Immersion}
  \left.
  \adjustbox{raise=2pt}{
  \begin{tikzcd}[
    row sep=-3pt,
    column sep=0pt
  ]
    \mathrm{d}\,
    \mathcolor{purple}{H_3}
      &=& 
    \phi^\ast G^s_4
    \\[+10pt]
    \mathrm{d}\, G_4
      &=&
    0
    \\
    \mathrm{d}\, G_7
    &=&
    G_4^s \, G_4^s
  \end{tikzcd}
  }
 \!\!\!\! \right\}
  \hspace{.4cm}
  \Leftrightarrow
  \hspace{.4cm}
  \left\{\!\!\!\!
  \scalebox{.9}{
    \begin{tabular}{c}
      $\sfrac{1}{2}$BPS immersion
      \\
      of {\color{purple}M5-worldvolume}
      \\
      in 11D SuGra solution.
    \end{tabular}
  }
  \right.
\end{equation}
As before, the most fundamental flux quantization law admissible for this system of Bianchi identities is 
\cite[\S 3.7]{FSS20-H}
a form of Cohomotopy, but now it is 3-Cohomotopy on the M5-worldvolume twisted by background 4-Cohomotopy, 
in that the classifying 3-spheres are fibered as the quaternionic Hopf fibration $S^3 \xrightarrow{\;} S^7 \xrightarrow{h_{\mathbb{H}}} S^4$:
\begin{equation}
  \label{Twisted3CohomotopyOnM5}
  \begin{tikzcd}[
    row sep=-2pt, 
    column sep=50pt
  ]
    \mathllap{
      \scalebox{.7}{
        \color{darkblue}
        \bf
        \def\arraystretch{.9}
        \begin{tabular}{c}
          Twisted 3-Cohomotopy
          \\
          on M5 worldvolume
        \end{tabular}
      }
    }
    \pi_0
    \,
    \mathrm{Maps}
    \left(
      \Sigma
      ,\,
      S^7
    \right)_{\!/S^4}
    \ar[
      rr,
      "{
        \scalebox{.7}{
          \color{darkgreen}
          \bf
          \def\arraystretch{.9}
          \begin{tabular}{c}
            character map in
            \\
            Hopf-twisted 3-Cohomotopy 
          \end{tabular}
        }
      }",
      "{
        \mathrm{ch}
      }"{swap}
    ]
    &&
    H^{1 + \phi^\ast G_4 
    }_{\mathrm{dR}}\big(
      X
      ;\,
      \mathfrak{l}_{{}_{S^4}}S^7
    \big)
    \\
    \mathllap{
      \scalebox{.7}{
        \color{darkblue}
        \bf
        \def\arraystretch{.9}
        \begin{tabular}{c}
        topological charge
        \\
        of B-field on M5
        \end{tabular}
      }
    }
    {[
      b
    ]}
    &\longmapsto&
    {\footnotesize
    \left[
    \def\arraycolsep{0pt}
    \def\arraystretch{1}
    \begin{array}{l}
      \mathrm{d}\, H_3\,=\,
      \phi^\ast G_4
    \end{array}
    \right]_{
      \mathrlap{
      \scalebox{.8}{$
      \left[
      \def\arraystretch{1}
      \def\arraycolsep{0pt}
      \begin{array}{l}
      \mathrm{d}\, G_7 \,=\,
      \tfrac{1}{2}G_4 \, G_4
      \\
      \mathrm{d}\, G_4 \,=\, 0
      \end{array}
      \right]
      $}
      }
    }
    }
  \end{tikzcd}
\end{equation}
(The notation on the right is for {\it twisted non-abelian} de Rham cohomology \cite[\S 6]{FSS23-Char} with coefficients in the Whitehead-$L_\infty$-algebra of the quaternionic Hopf fibration, but what this means here is just exactly the classes of fluxes satisfying Bianchi identities as shown on the botton right.)

And again, considering this twisted cohomotopical flux quantization on the M5-brane provably implies some subtle topological effects expected for M5-branes (cf. \cite{FSS21-Hopf}\cite{FSS21-TwistedString}), and hence we assume this here.

\medskip 
But we now take into account one more field on the worldvolume, to enhance this situation a little further:

\medskip

\noindent
{\bf Chern-Simons A-field on M5.} In view of this effective re-definition --- of on-shell 11D supergravity with probe branes --- in terms of (non-linear) Bianchi identities for super-flux densities on
super-torsion-free super-space, we may go ahead and consider a further super-flux density
$$
  F_2^s
  \;\;
  \defneq
  \;\;
  (F_2)_{a_1 a_2}
  \,
  e^{a_1}\, e^{a_2}
$$
on the M5's super-worldvolume with super-coframe $(e,\psi)$, subjected to the Bianchi identity for an ordinary gauge field, but again imposed on super-space:
$$
  \mathrm{d}
  \,
  F_2^s
  \;\;
  =
  \;\;
  0
  \;\;\;\;\;\;\;
  \Leftrightarrow
  \;\;\;\;\;\;\;
  \left\{\!\!
  \def\arraystretch{1.5}
  \begin{array}{l}
    \big(
    \nabla_{a_1}
    (F_2)_{a_2 a_3}
    \big)
    e^{a_1} e^{a_2} e^{a_3}
    \;=\;
    0
    \\
    \big((F_2)_{a b}\big)
    e^a \big(
      \delta^b_{b'}
      -2(\tilde H^2_3)^b_{b'}
    \big)
    \big(\hspace{1pt}
      \overline{\psi}
      \,\gamma^{b'}\,
      \psi
    \big)
    \;=\;
    0
  \end{array}
 \!\! \right\}
  \;\;\;\;\;\;\;
  \Leftrightarrow
  \;\;\;\;\;\;\;
  F_2 \;=\; 0
  \,.
$$
The super-coframe components of this Bianchi identity shown in the middle -- computed via \cite[(122)]{GSS24-FluxOnM5} --,
immediately show that this is equivalent simply to adjoining the further equation of motion shown on the right, 
which is that of an abelian Chern-Simons gauge field (not back-reacting on the M5/SuGra background). 
Since $F^s_2$ thus vanishes on-shell, we may now consider adding its square to the Bianchi identity for $H^s_3$ without actually changing its on-shell content:
\begin{equation}
  \label{GaussLawOnM5ImmersionWithCS}
  \left.
  \adjustbox{raise=2pt}{
  \begin{tikzcd}[
    row sep=-4pt,
    column sep=0pt
  ]
    \mathrm{d}\,
    \mathcolor{purple}{F_2^s} &=& 0
    \\
    \mathrm{d}\,
    H^s_3 
      &=& 
    \phi^\ast G^s_4
    \\
    && 
    \mathcolor{purple}{+ F_2^s\, F_2^s}
    \\[+4pt]
    \mathrm{d}\, G^s_4
      &=&
    0
    \\
    \mathrm{d}\, G^s_7
    &=&
    \tfrac{1}{2} G_4^s \, G_4^s
  \end{tikzcd}
  }
 \!\!\!\! \right\}
  \hspace{.4cm}
  \Leftrightarrow
  \hspace{.4cm}
  \left\{\!\!\!\!
  \scalebox{.9}{
    \begin{tabular}{c}
      $\sfrac{1}{2}$BPS immersion
      \\
      of M5-worldvolume
      \\
      \color{purple}
      with CS gauge field
      \\
      in 11D SuGra solution
    \end{tabular}
  }
  \right.
\end{equation}

While the CS field thus contributes nothing, on-shell, to the flux densities and their equations of motion, hence is invisible to local on-shell analysis,
it does affect the {\it torsion content} of the theory, in that the admissible flux quantization laws change: The above \eqref{GaussLawOnM5ImmersionWithCS} is no longer flux-quantized by the 3-spheres that arrange to the
quaternionic Hopf fibration $S^3 \xrightarrow{} S^7 \xrightarrow{h_{\mathbb{H}}} S^4$ but it does turn out \cite{FSS22-Twistorial}\cite{SS25-EquTwistorial}
to be flux-quantized by 2-spheres that arrange into the ``Atiyah-Penrose twistor fibration'' $S^2 \xrightarrow{} \mathbb{C}P^3 \xrightarrow{t_{\mathbb{H}}} S^4$, classifying {\it twistorial Cohomotopy} (see also \cite[Ex. 3.11]{FSS23-Char}\cite{SS23-MF}):
\begin{equation}
  \label{CharacterInTwistorialCohomotopy}
  \hspace{-2cm}
  \begin{tikzcd}[
    row sep=-1pt, 
    column sep=45pt
  ]
    \pi_0
    \,
    \mathrm{Maps}
    \left(
      \Sigma
      ,\,
      \mathbb{C}P^3
    \right)_{\!/S^4}
    \ar[
      rr,
      "{
        \scalebox{.7}{
          \color{darkgreen}
          \bf
          \def\arraystretch{.9}
          \begin{tabular}{c}
            character map in
            \\
            Twistorial Cohomotopy
          \end{tabular}
        }
      }",
      "{
        \mathrm{ch}
      }"{swap}
    ]
    &&
    H^{1 + \phi^\ast G_4 
    }_{\mathrm{dR}}\big(
      X
      ;\,
      \mathfrak{l}_{{}_{S^4}}
      \mathbb{CP}^3
    \big)
    \\
    \mathllap{
      \scalebox{.7}{
        \color{darkblue}
        \bf
        \def\arraystretch{.9}
        \begin{tabular}{c}
          topological charge
          \\
          of A-field on M5 
        \end{tabular}
      }
    }
    {[
      a
    ]}
    &\longmapsto&
    {\footnotesize
    \left[
    \def\arraycolsep{0pt}
    \def\arraystretch{1}
    \begin{array}{l}
      \mathrm{d}\, F_2 \,=\, 0
      \\
      \mathrm{d}\, H_3\,=\,
      \phi^\ast G_4 + 
      F_2 \, F_2
    \end{array}
    \right]_{
      \mathrlap{
      \scalebox{.8}{$
      \left[
      \def\arraystretch{1}
      \def\arraycolsep{0pt}
      \begin{array}{l}
      \mathrm{d}\, G_7 \,=\,
      \tfrac{1}{2}G_4 \, G_4
      \\
      \mathrm{d}\, G_4 \,=\, 0
      \end{array}
      \right]
      $}
      }
    }
}
  \end{tikzcd}
\end{equation}

This is interesting for our purpose, because {\it when} the pullback of the background C-field flux to the M5 worldvolume vanishes, then twistorial Cohomotopy reduces to plain 2-Cohomotopy and anyonic solitons appear \cite{SS25-TQBits}.

$$
  \begin{tabular}{p{7.7cm}}
    \footnotesize
    {\bf Flux quantization of the M5's A/B-field} in twistorial Cohomotopy means that its topological sectors are classified by maps from the worldvolume $\Sigma$ to $\mathbb{C}P^3$ which lift the classifying map 
    \eqref{CohomotopyCharacter}
    of the bulk C-field
    through the twistor fibration $t_{\mathbb{H}}$. 
    In the case that the bulk C-field trivializes on the M5, as shown at the bottom, this means that the A/B-field sector is equivalently classified by maps to a single $S^2$-fiber of the twistor fibration, hence is flux-quantized in plain 2-Cohomotopy.
  \end{tabular}
  \hspace{1cm}
  \adjustbox{raise=10pt}{
  \begin{tikzcd}[column sep=large]
    & 
    S^2
    \ar[d]
    \ar[r]
    &
    \mathbb{C}P^3
    \ar[
      d,
      "{
        t_{\mathbb{H}}
      }"
    ]
    \\
    \Sigma
    \ar[r]
    \ar[
      ur,
      dashed,
      "{ a }"
    ]
    \ar[
      rr,
      bend right=25pt,
      "{
        \phi^\ast c
      }"{swap},
      "{ \Uparrow }"
    ]
    &
    \ast
    \ar[r]
    &
    S^4
  \end{tikzcd}
  }
$$

However, for realizing anyons we need this situation not on all of the worldvolume $\Sigma$, but just on a 3D submanifold.
What we observe next is that this vanishing of the background twist by the C-field on a sub-worldvolume
may itself be enforced by further refining the flux quantization law to {\it equivariant} cohomology measuring charges on orbifolds.


\section{Twistorial Quantization on Orbi-M5s}
\label{TwistorialFluxQuantization}

We finally generalize the previous discussion to orbifolded M5-worldvolumes. This requires passing to {\it equivariant} generalized cohomology.

\smallskip

\noindent
{\bf Equivariant generalized cohomology.}
With generalized cohomology understood as mapping classes into classifying spaces (per \S\ref{SolitonsViaFluxQuantization}), it is clear what to make of {\it equivariant} cohomology on $G$-orbifolds $X \!\sslash\! G$:
\footnote{
  We take the equivariance group $G$ here to be a finite group. In fact, in the application below, we consider just $G \defneq \ZTwo$.
}
This must be given by homotopy classes of $G$-equivariant maps into classifying spaces 
$G \acts \, \mathcal{A}$ equipped with continuous $G$-actions (math jargon: ``$G$-spaces''):
\vspace{-1mm} 
\begin{equation}
  \label{EquivariantCohomology}
  H^1_G(X;\, \Omega\mathcal{A})
  \;\;
  :=
  \;\;
  \pi_0
  \,
  \mathrm{Maps}\big(
    X
    ,\,
    \mathcal{A}
  \big)^G
  \;\;
  =
  \;\;
  \bigg\{\!\!\!\!
  \adjustbox{
    raise=-6pt
  }{
  \begin{tikzcd}
    X
    \ar[
      in=60,
      out=180-60,
      looseness=3.5,
      shift right=3pt,
      "{
        G
      }"{description}
    ]
    \ar[
      rr,
      "{ \Phi }"
    ]
    &
    &
    \mathcal{A}
    \ar[
      in=60,
      out=180-60,
      looseness=3.5,
      shift right=3pt,
      "{
        G
      }"{description}
    ]
  \end{tikzcd}
  }
  \!\!\!\bigg\}_{\!\!\big/\mathrm{hmtp}}
  .
\end{equation}
Indeed, the equivariant refinement of topological K-theory, quantizing RR-flux (and hence fractional D-brane charge) on orbifolds 
\cite{SzaboValentino10}
is of this form \cite[Ex. 4.5.4]{SS21-EBund}.

A subtle effect of this innocent-looking definition \eqref{EquivariantCohomology} -- rooted in {\it Elmendorf's theorem}, cf. \cite[Prop. 4.5.1]{SS21-EBund} -- is that equivariant homotopy theory is like ordinary homotopy theory applied to the system \footnote{
  Technically, this ``system'' of fixed loci is a presheaf of spaces on the orbit category of $G$, cf. \cite[Ex. 2.20]{SS25-EquTwistorial}
}
of $H$-fixed loci $X^H$ for all subgroups $H \subset G$:
$$
  X^H 
    \;:=\;
  \Big\{
    x \,\in\, X
    \;\big\vert\;
    \underset{h \in H}{\forall}
    \;\;\;
      h \cdot x \,=\, x
  \Big\}
  \;\;
  \subset
  \;\;
  X
  \,.
$$
Here we only need to know that, in particular, the character map on any $\ZTwo$-equivariant $\mathcal{A}$-cohomology takes values 
\cite[\S 3]{SS25-EquTwistorial}
in {\it pairs} of Bianchi identities, one on $X \,=\, X^{\{\mathrm{e}\}}$ and one on its fixed locus $X^{\ZTwo}$:
\vspace{-1mm} 
\begin{equation}
  \label{EquivariantBianchiIdentities}
  \begin{tikzcd}[
    row sep=-3pt,
    ampersand replacement=\&
  ]
    H^1_{\ZTwo}\big(
      X
      ;\,
      \Omega\mathcal{A}
    \big)
    \;:=\;
    \pi_0
    \, 
    \mathrm{Maps}\big(
      X
      ,\,
      \mathcal{A}
    \big)^{\ZTwo}
    \ar[
      rr,
      "{
        \scalebox{.7}{
          \color{darkgreen}
          \bf
          \begin{tabular}{c}
            character map in
            \\
            equivariant \scalebox{1.2}{$\mathcal{A}$}-cohomology
          \end{tabular}
        }
      }",
      "{
        \mathrm{ch}
      }"{swap}
    ]
    \&\&
    H^1_{\mathrm{dR},\,\ZTwo}\big(
      X;\, \mathfrak{l}\mathcal{A}
    \big)
    \\
    \mathllap{
      \scalebox{.7}{
        \color{darkblue}
        \bf
        \def\arraystretch{.9}
        \begin{tabular}{c}
          topological charge
          \\
          on $\ZTwo$-orbifold
        \end{tabular}
      }
    }
    {[\rchi]}
    \& \longmapsto\&
    \hspace{2cm} 
    \left\{\!\!\!\! \footnotesize
    \def\arraystretch{1.3}
    \begin{array}{ccc}
      \big[
        \mathrm{d}\, F^{(i)} \;=\;
        P^i\big(\vec F\,\big)
      \big]
      &&
      X
      \\
      \rotatebox[origin=c]{-90}{$\mapsto$}\
      \,
      \scalebox{.9}{$\iota^\ast$}
      &
      \scalebox{.8}{on}
      &
      \rotatebox[origin=c]{90}{$\hookrightarrow$}
      \mathrlap{\,\scalebox{.9}{$\iota$}}
      \\
      \big[
        \mathrm{d}\, f^{(i)} \;=\;
        p^i\big(\vec f\;\big)
      \big]
      &&
      X^{\ZTwo}
    \end{array}
    \right.
  \end{tikzcd}
\end{equation}

This may be understood from physics, where on orbifolded super-space the Bianchi identities may simplify as one restricts to orbi-singularities:

\medskip

\noindent
{\bf The super-orbifolding.} Consider the super-tangent space  $\mathbb{R}^{1,5\,\vert\,2 \cdot \mathbf{8}_+}$ 
of an M5 super-worldvolume at the locus of a $\ZTwo$-orbisingularity
via rotation by the angle $\pi$ in the $(4,5)$-plane, as indicated by the last factor in the brane diagram \eqref{TheBraneDiagram}, hence by acting with the super-isometry
$$
  \exp\big(
     \pi
     \,
     \tfrac{1}{2} \gamma_{45}
  \big)
  \;\;
  =
  \;\;
  \gamma_{45}
  \;\;
  \in
  \;\;
  \mathrm{Spin}(1,5)
  \;\subset\;
  \mathrm{Iso}\big(
    \mathbb{R}^{1,5\,\vert\,2\cdot \mathbf{8}_+}
  \big)
  \,.
$$
This generates a cyclic group of order 4, being the spin double cover of the ordinary $\pi$-rotation action of $\mathbb{Z}_2 \,\subset\, \mathrm{SO}(1,5)$, whence we suggestively denote it by $\widehat{\ZTwo}$:
$$
  \begin{tikzcd}[row sep=10pt]
    \{\pm 1\}
    \ar[d, hook]
    \ar[r, equals]
    &
    \{\pm 1\}
    \ar[d, hook]
    \\
    \mathllap{
      \langle 
        \gamma_{45}
      \rangle
      \;\simeq\;
      \mathbb{Z}_4
      \;=:\;\;\;
    }
    \widehat{\ZTwo}
    \ar[d, ->>]
    \ar[r, hook]
    &
    \mathrm{Spin}(1,5)
    \ar[
      d,
      ->>
    ]
    \\
    \ZTwo
    \ar[
      r,
      hook
    ]
    &
    \mathrm{SO}(1,5)
    \mathrlap{\,.}
  \end{tikzcd}
$$
Since $\gamma_{45} \gamma_{45} \,=\, - 1$, 
this action fixes none of the spinors except $0 \,\in\, 2 \cdot \mathbf{8}_+$, and hence its fixed super-locus is just bosonic 4D spacetime:
\footnote{
Besides the full fixed super-locus \eqref{SuperFixedLocus}, there is also the intermediate one for the unique non-trivial proper subgroup $\{\pm 1\}\subset\, \widehat{\ZTwo}$ of order 2, which is
$
  \big(
    \mathbb{R}^{1,5\,\vert\,2 \cdot \mathbf{8}_=}
  \big)^{\{\pm 1\}}
  \;=\;
  \mathbb{R}^{1,5\,\vert\, \mathbf{0}}
$.
But since this locus does not further add to the discussion, for notational brevity we disregard it in the following.
}
\begin{equation}
  \label{SuperFixedLocus}
  \big(
    \mathbb{R}^{1,6\,\vert\,2 \cdot \mathbf{8}_+}
  \big)^{\widehat{\ZTwo}}
  \;=\;
  \mathbb{R}^{1,3\,\vert\,\mathbf{0}}
  \xhookrightarrow{\quad \iota \quad}
  \mathbb{R}^{1,5\,\vert\,2 \cdot \mathbf{8}_+}
  \,.
\end{equation}

\medskip

\noindent
{\bf The A/B-field on orbi-M5s.}
Since therefore the avatar $(G_4^0, G_7^0)$ of the C-field super-flux densities
\eqref{11DSuperFluxDensities}
necessarily vanishes on the fixed locus, $\iota^\ast (G_4^0,\,G_7^0) \,=\, 0$, it is consistent to demand that in fact $\iota^\ast (G_4^s,\,G_7^s) \,=\, 0$ and hence to ask the
$\ZTwo$-equivariant enhancement -- according to  \eqref{EquivariantBianchiIdentities} -- of the previous Bianchi identities \eqref{GaussLawOnM5ImmersionWithCS} to be of  form shown on the right in \eqref{EquivariantTwistorialCharacter} below. 
Now the result of \cite{SS25-EquTwistorial} says that this is indeed the character image of a $\ZTwo$-equivariantization 
\eqref{EquivariantCohomology}
of the above twistorial Cohomotopy \eqref{CharacterInTwistorialCohomotopy},
namely of {\it equivariant twistorial Cohomotopy} classified by $\ZTwo \acts \, \mathbb{C}P^3$ where $\ZTwo$ acts by permuting the two factors in 
$$
  \mathbb{C}P^3 \,\simeq\, 
\big( 
  (
  \mathbb{C}^2
  \times
  \mathbb{C}^2
  ) \setminus \{0\}
\big)\big/\mathbb{C^\times}
\,,
$$
so that its fixed locus is the 2-sphere
\begin{equation}
  \label{FixedLocusInCP3}
  (\mathbb{C}P^3)^{\ZTwo}
  \;\simeq\;
  \big(
    \mathbb{C}^2 \setminus \{0\}
  \big)/\mathbb{C}^\times
  \;\simeq\;
  \mathbb{C}P^1
  \;\simeq\;
  S^2
  \,.
\end{equation}
Namely, we have from \cite[Thm. 1.1]{SS25-EquTwistorial}
\begin{equation}
  \label{EquivariantTwistorialCharacter}
  \hspace{3mm} 
  \begin{tikzcd}[
    sep=0pt,
    ampersand replacement=\&
  ]
  \mathllap{
    \scalebox{.7}{
      \color{darkblue}
      \bf
      \def\arraystretch{.9}
      \begin{tabular}{c}
        Equivariant 
        \\
        twistorial
        \\
        Cohomotopy
      \end{tabular}
    }
  }
  \pi_0
  \,
  \mathrm{Maps}\big(
    \Sigma
    ,\,
    \mathbb{C}P^3
  \big)_{\!/S^4}^{\ZTwo}
  \ar[
    rr,
    "{
      \scalebox{.7}{
        \color{darkgreen}
        \bf
        \def\arraystretch{.9}
        \begin{tabular}{c}
          character map in equivariant
          \\
          twistorial Cohomotopy
        \end{tabular}
      }
    }",
    "{
      \mathrm{ch}
    }"{swap}
  ]
  \&\phantom{------}\&
  H^1_{\mathrm{dR},\, \ZTwo}\big(
    \Sigma
    ;\,
    \mathfrak{l}_{S^4}
    \mathbb{C}P^3
  \big)
  \&\&
  \\[-5pt]
  {}
  \&\&
  \left[\!\! \footnotesize 
  \def\arraystretch{1}
  \begin{array}{ccl}
    \mathrm{d}
    \, F_2
    &=&
    0
    \\
    \mathrm{d}\,H_3
    &=&
    \phi^\ast G_4
    \,+\, F_2 \, F_2
    \\
    \mathrm{d}\, G_4 &=& 0
    \\
    \mathrm{d}\, G_7 &=& 
    \tfrac{1}{2} G_4 \, G_4
  \end{array}
  \!\!\right]
  \&\&
  \Sigma
  \\
  \mathllap{
    \scalebox{.7}{
      \color{darkblue}
      \bf
      \def\arraystretch{.9}
      \begin{tabular}{c}
        topological charge
        \\
        of A/B-field on
        \\
        {\color{purple}orbifolded}
        M5
      \end{tabular}
    }
  }
  {[a]}
  \&
  \hspace{1cm} \longmapsto \hspace{2cm}
  \&
  \hspace{2cm} 
  \rotatebox[origin=c]{-90}{$\longmapsto$}
  \, 
  \scalebox{.8}{$\iota^\ast$}
  \&
  \mbox{on}
  \\
  {}
  \&\&
  \left[\!\!  \footnotesize
  \def\arraystretch{1}
  \begin{array}{ccl}
    \mathrm{d}\, F_2 &=& 0
    \\
    \mathrm{d}\, H_3 &=& F_2 \, F_2
  \end{array}
  \!\!\right]
  \&\&
  \Sigma^{\ZTwo}
  \ar[
    uu,
    hook,
    "{ \iota }"
  ]
  \end{tikzcd}
\end{equation}.

\vspace{-1mm} 
Such Green-Schwarz-type Bianchi identities on M5-branes (like the $\mathrm{d}\, H_3 \,=\, F_2 \, F_2$ appearing here)  are traditionally motivated by anomaly cancellation, cf. \cite[(1.32)]{CDI21}.

\medskip

\noindent
{\bf Computing charge in equivariant twistorial Cohomotopy.}
For M5-brane orbi-topologies of interest here \eqref{TheBraneDiagram}, with worldvolume
\begin{equation}
  \label{OrbiWOrldvolume}
  \ZTwo\acts \; \Sigma^{1,5}
  \;=\;\,
  \Sigma^{1,3} 
    \,\wedge\,
  (\ZTwo \acts \; \mathbb{R}^2_{\plus})
\end{equation}
the inclusion of the fixed locus is evidently a $\ZTwo$-equivariant homotopy equivalence
\begin{equation}
  \label{M5FixedLocusInclusionIsEquivalence}
  (\Sigma^{1,5})^{\ZTwo}
  \,=\,
  \Sigma^{1,3}
\xhookrightarrow[\ZTwo]{\quad\sim\quad}
  \Sigma^{1,6}
  \,.
\end{equation}
This implies that homotopy-classes of $\ZTwo$-equivariant maps from $\Sigma$ to $\mathbb{C}P^3$ are in bijection to homotopy classes of plain maps between the fixed loci --- but this is just the 2-Cohomotopy of the M5s orbi-singularity! Namely:
\begin{equation}
  \label{ETCreducingToPlainCohomotopy}
  \overset{
    \mathclap{
      \adjustbox{
        scale=.7,
        raise=2pt
      }{
        \color{darkblue}
        \bf
        \def\arraystretch{.9}
        \begin{tabular}{c}
          equivariant 
          \\
          twistorial
          Cohomotopy
        \end{tabular}
      }
    }
  }{
  \pi_0
  \,
  \mathrm{Maps}\big(
    \underset{
      \mathclap{
        \adjustbox{
          raise=-9pt,
          scale=.7
        }{
          \color{darkblue}
          \bf
          \def\arraystretch{.9}
          \begin{tabular}{c}
            of orbi-
            \\
            worldvolume
          \end{tabular}
        }
      }
    }{
      \Sigma^{1,5}
    }
    ,\,
    \mathbb{C}P^3
  \big)_{\!/S^4}^{\ZTwo}
  }
  \;\;
  \underset{
    \mathclap{
      \adjustbox{
        scale=.7,
        raise=-2pt
      }{
        \color{gray}
        \eqref{M5FixedLocusInclusionIsEquivalence}
      }
    }
  }{\simeq}
  \;\;
  \pi_0
  \,
  \mathrm{Maps}\Big(
    (\Sigma^{1,5})^{\ZTwo}
    ,\,
    (\mathbb{C}P^3)^{\ZTwo}
  \Big)
  \;\;
  \underset{
    \mathclap{
      \adjustbox{
        scale=.7,
        raise=-2pt
      }{
        \color{gray}
        \eqref{FixedLocusInCP3}
      }
    }
  }{\simeq}
  \;\;
  \pi_0
  \,
  \mathrm{Maps}\big(
    \Sigma^{1,3}
    ,\,
    S^2
  \big)
  \;\;
  \defneq
  \;\;
  \overset{
    \mathclap{
      \adjustbox{
        scale=.7,
        raise=2pt
      }{
        \color{darkblue}
        \bf
        \def\arraystretch{.9}
        \begin{tabular}{c}
          plain
          \\
          2-Cohomotopy
        \end{tabular}
      }
    }
  }{
  \pi^2\big(
    \underset{
      \mathclap{
        \adjustbox{
          scale=.7,
          raise=-8pt
        }{
          \color{darkblue}
          \bf
          \def\arraystretch{.9}
          \begin{tabular}{c}
            of orbi-
            \\
            singularity
          \end{tabular}
        }
      }
    }{
      \Sigma^{1,3}
    }
  \big)
  }
  \,.
\end{equation}

In conclusion, so far:
\vspace{-2mm} 
\begin{center}
\colorbox{lightgray}{\parbox{0.77\textwidth}{
  \it
  Flux quantization on orbifolded M5-worldvolumes in equivariant twistorial Cohomotopy,
  
  which generally reflects a list of expected M-theoretic topological effects, 
  
  specializes on orbi-M5 worldvolumes of the form \eqref{TheBraneDiagram} 
  
  to flux quantization in 2-Cohomotopy of the orbisingularity $\Sigma^{1,3} \subset \Sigma^{1,5}$.
  }}
\end{center}

\vspace{-1mm} 
But this is most interesting, because codimension=2 solitons flux-quantized in 2-Cohomotopy are anyonic \cite{SS24-AbAnyons}:


\vspace{1mm} 
\noindent
{\bf Soliton moduli on Seifert orbi-M5s.}
Finally we fully specialize the M5-worldvolume domain $\Sigma^{1,5}$ from \eqref{OrbiWOrldvolume} to the case announced in \eqref{TheBraneDiagram}, where:
$$
  \underset{
    \mathclap{
      \adjustbox{
        rotate=-12,
        scale=.7
      }{
        \hspace{-5pt}
        \rlap{
        \color{darkblue}
        \bf
          M5 orbi-worldvolume
        }
      }
    }
  }{
  \ZTwo \acts \;
  \Sigma^{1,5}
  }
  \;\;:=\;\;
  \underset{
    \mathclap{
      \adjustbox{
        rotate=-12,
        scale=.7
      }{
        \hspace{-5pt}
        \rlap{
        \color{darkblue}
        \bf
          time axis
        }
      }
    }
  }{
    \mathbb{R}^{1,0}_{\plus}
  }
  \,\wedge\,
  \underset{
    \mathclap{
      \adjustbox{
        rotate=-12,
        scale=.7
      }{
        \hspace{-5pt}
        \rlap{
        \color{darkblue}
        \bf
          transverse space
        }
      }
    }
  }{
  \mathbb{R}^2_{\cpt}
  }
  \,\wedge\,
  \underset{
    \mathclap{
      \adjustbox{
        rotate=-12,
        scale=.7
      }{
        \hspace{-5pt}
        \rlap{
        \color{darkblue}
        \bf
          M-theory circle
        }
      }
    }
  }{
  \mathbb{R}^1_{\cpt}
  }
  \,\wedge\,
  \underset{
    \mathclap{
      \adjustbox{
        rotate=-12,
        scale=.7
      }{
        \hspace{-5pt}
        \rlap{
        \color{darkblue}
        \bf
          2D orbifold
        }
      }
    }
  }{
  (\ZTwo \acts \mathbb{R}^2_{\plus})
  }
  \,.
$$

\vspace{.5cm}
\newpage 

\noindent
(Here we are considering the ``uncompactified'' M-theory circle, $\mathbb{R}^1$, as befits the fully non-perturbative situation, but we retain the condition that solitonic fields vanish-at-infinity also in this direction -- which still gives an effective circle but with reachable point-at-infinity.)
It follows that the topological quantum observables in degree=0 --- as in \eqref{ElectromagneticObservables} --- on such flux-quantized M5-worldvolume domains are: 
\begin{equation}
  \label{QuantumObservablesOnM5}
  \def\arraystretch{1.5}
  \begin{array}{ccl}
  \mathrm{Obs}_0
  &:=&
    \overset{
      \mathclap{
        \adjustbox{
          raise=13pt,
          scale=.7
        }{
          \color{darkblue}
          \bf
          \def\arraystretch{.9}
          \begin{tabular}{c}
            homology\;\;\;\;\;
            \\ 
            of
          \end{tabular}
        }
      }
    }{
  H_0
  }
  \Big(
    \overset{
      \mathclap{
        \scalebox{.7}{
          \color{darkblue}
          \bf
          \def\arraystretch{.9}
          \begin{tabular}{c}
            moduli of equivariant 
            \\ 
            twistorial Cohomotopy
          \end{tabular}
        }
      }
    }{
  \mathrm{Maps}\big(
    \Sigma^{1,5}
    ,\,
    \mathbb{C}P^3
  \big)_{\!/S^4}^{\ZTwo}
  }
  ;\,
  \mathbb{C}
  \Big)
  \,
  \underset{
    \mathclap{
    \adjustbox{
      raise=-3pt,
      scale=.7
    }
    {
      \eqref{ETCreducingToPlainCohomotopy}
    }
    }
  }{
    \simeq
  }
  \,
  H_0\Big(
    \overset{
      \mathclap{
        \scalebox{.7}{
          \color{darkblue}
          \bf
          \def\arraystretch{.9}
          \begin{tabular}{c}
          \end{tabular}
        }
      }
    }{
    \mathrm{Maps}\big(
      \mathbb{R}^{2}_{\cpt}
      \,\wedge\,
      \mathbb{R}^1_{\cpt}
      ,\,
      S^2
    \big)
    }
    ;\,
    \mathbb{C}
  \Big)
  \\
  &
  \underset{
    \mathclap{
    \adjustbox{
      raise=-2pt,
      scale=.7
    }{
      \eqref{SmashOfSpheres}
    }
    }
  }{
  \simeq
  }
  &
  H_0\Big(
    \Omega
    \,
    \grayunderbrace{
    \mathrm{Maps}\big(
      \mathbb{R}^{2}_{\cpt}
      ,\,
      S^2
    \big)
    }{
      \mathbb{G}
      \mathrm{Conf}(\mathbb{R}^2)
    }
    ;\,
    \mathbb{C}
  \Big)
  \,.
  \end{array}
\end{equation}
Remarkably, as indicated under the brace, a classical theorem \cite{Segal73} says that 
$
  \mathrm{Maps}\big(
    \mathbb{R}^2_{\cpt}
    ,\,
    S^2
  \big)
  \;
  \simeq
  \;
  \mathbb{G}
  \big(
  \mathrm{Conf}(\mathbb{R}^2)
  \big)
$
is the ``group-completed configuration space of points'' on the plane, which means that:

\begin{itemize}[
  itemsep=2pt,
  topsep=2pt
]
\item[\bf (i)]
$\mathrm{Conf}(\mathbb{R}^2)$ is the space of finite sub-sets of $\mathbb{R}^2$ (cf. \cite{Kallel24}), here to be thought of as configurations of  unit-charged solitons in their transverse space,

\item[\bf (ii)]
$\mathbb{G}(\cdots)$ is the analogous space given by including also {\it anti-solitons} of negative unit charge -- so that their total number forms the additive group $\mathbb{Z}$ under pair creation/annihilation, instead of just the semi-group $\mathbb{N}$ (whence ``group-completion'');

\item[\bf (iii)]
$\Omega (\cdots)$ is the space of based loops of such soliton/anti-soliton configurations, hence of trajectories where soliton/anti-soliton pairs appear out of the vacuum by pair creations, then braid their worldlines around each other, and finally annihilate again into the vacuum, thus forming {\it links} in $\mathbb{R}^3$ (as indicated in \hyperlink{SolitonsBraiding}{\it Figure B});

\item[\bf (iv)]
more precisely, \cite{Okuyama05} says that $\mathbb{G}(-)$ is a regularized form of the naive soliton/anti-soliton configuration space, which by \cite{SS24-AbAnyons} means that $\Omega \, \mathbb{G}(-)$ is not just the naive space of links but the space of {\it framed links}, where the framing is just that traditionally used to regularize Wilson-line observables of Chern-Simons theory!
\end{itemize}

From here on, the analysis of the quantum states of these solitons proceeds just as in \cite{SS24-AbAnyons}. \footnote{
  In \cite{SS24-AbAnyons}, the passage from the twisted 3-Cohomotopy 
  \eqref{Twisted3CohomotopyOnM5}
  on M5 worldvolumes to plain 2-Cohomotopy was enforced by orientifolding 
  after assuming the background C-field charge to vanish. But since it is not so clear how to orientifold the twisted 3-Cohomotopy in the presence of background charge we are here improving on this model by instead flux-quantizing in twistorial Cohomotopy on orbifolded worldvolumes.
}

\hspace{-2mm}
\begin{tabular}{cc}
\hypertarget{SolitonsBraiding}{}
\begin{tabular}{p{10.8cm}}
  \footnotesize
  {\bf Figure B.}
  A generic element in the moduli space of solitons on M5 orbi-worldvolumes flux-quantized in equivariant twistorial Cohomotopy is, according to \eqref{QuantumObservablesOnM5}, a process where soliton/anti-soliton pairs in the first factor of \eqref{TheBraneDiagram}
  appear out of the vacuum, braid their worldlines around each other, and finally annihilate into the vacuum again. 

\smallskip 

  Just such processes are traditionally considered as topological quantum computation protocols (cf.
 \cite[Fig. 17]{Kauffman02}\cite[Fig. 2]{FKLW03}\cite[Fig. 2]{Rowell16}\cite[Fig. 2]{DMNW17}\cite[Fig. 3]{RowellWang18}\cite[Fig. 1]{Rowell22}
 \cite[Fig. 7]{SV25}).

 \smallskip

  The close analysis of \cite[\S 3]{SS24-AbAnyons} shows that the resulting {\it links} are in fact {\it framed}, which is known as the regularization of the corresponding Chern-Simons Wilson loop observables.
\end{tabular}
&
\adjustbox{
  raise=-2.4cm,
  scale=.8
}{
\begin{tikzpicture}

\begin{scope}[xscale=-1.1]
\draw[
  line width=.27cm,
  gray,
]
  (0,0) 
    .. controls (0,1) and (1.7,1) .. 
  (1.7,2)
    .. controls (1.7,3) and (0,3) ..
  (0,4);
\draw[
  line width=.22cm,
  white,
]
  (0,0) 
    .. controls (0,1) and (1.7,1) .. 
  (1.7,2)
    .. controls (1.7,3) and (0,3) ..
  (0,4);
\end{scope}
\draw[line width=.25cm]
  (0,0) 
    .. controls (0,1) and (1.7,1) .. 
  (1.7,2)
    .. controls (1.7,3) and (0,3) ..
  (0,4);
\begin{scope}[xscale=-1]
\draw[
  line width=.22cm,
  white,
  draw opacity=.5
]
  (0,0) 
    .. controls (0,1) and (1.7,1) .. 
  (1.7,2)
    .. controls (1.7,3) and (0,3) ..
  (0,4);
\end{scope}


\begin{scope}[
  shift={(2,0)},
  xscale=-1
]
\draw[line width=.4cm, white]
  (0,0) 
    .. controls (0,1) and (1.7,1) .. 
  (1.7,2);
\begin{scope}[xscale=-1]
\draw[
  line width=.27cm,
  gray,
]
  (0,0) 
    .. controls (0,1) and (1.7,1) .. 
  (1.7,2)
    .. controls (1.7,3) and (0,3) ..
  (0,4);
\draw[
  line width=.22cm,
  white,
]
  (0,0) 
    .. controls (0,1) and (1.7,1) .. 
  (1.7,2)
    .. controls (1.7,3) and (0,3) ..
  (0,4);
\end{scope}
\draw[line width=.25cm]
  (0,0) 
    .. controls (0,1) and (1.7,1) .. 
  (1.7,2)
    .. controls (1.7,3) and (0,3) ..
  (0,4);
\begin{scope}[xscale=-1]
\draw[
  line width=.22cm,
  white,
  draw opacity=.5
]
  (0,0) 
    .. controls (0,1) and (1.7,1) .. 
  (1.7,2)
    .. controls (1.7,3) and (0,3) ..
  (0,4);
\end{scope}
\end{scope}


\begin{scope}
\clip
  (0,2.5)
  rectangle (2,3.3);

\draw[line width=.4cm, white]
  (0,0) 
    .. controls (0,1) and (1.7,1) .. 
  (1.7,2)
    .. controls (1.7,3) and (0,3) ..
  (0,4);
\draw[line width=.25cm]
  (0,0) 
    .. controls (0,1) and (1.7,1) .. 
  (1.7,2)
    .. controls (1.7,3) and (0,3) ..
  (0,4);
\end{scope}

\node at (0,-.25) {$\varnothing$};
\node at (2,-.25) {$\varnothing$};
\node at (0,4.2) {$\varnothing$};
\node at (2,4.2) {$\varnothing$};

\end{tikzpicture}
}

\end{tabular}

\medskip

\noindent
{\bf Topological-sector light-cone quantization.} 
The homology of a loop space as in \eqref{QuantumObservablesOnM5} carries the extra structure of a star-algebra (in fact of a Hopf algebra), known as the {\it Pontrjagin algebra}-structure, whose product operation is the push-forward in homology of the 
operation ``$\mathrm{conc}$''
of concatenation of loops (cf. \cite[(20)]{SS23-Obs}): 
$$
  \begin{tikzcd}[
    column sep=20pt,
    row sep=0pt
  ]
  H_\bullet\big(
    \Omega \, \mathrm{Fields}
    ;\,
    \mathbb{C}
  \big)
  \otimes
  H_\bullet\big(
    \Omega \, \mathrm{Fields}
    ;\,
    \mathbb{C}
  \big)
  \ar[
    r,
    phantom,
    "{ \simeq }",
    "{
      \scalebox{.6}{
        \color{darkgreen}
        \bf
        \def\arraystretch{.9}
        \begin{tabular}{c}
          K{\"u}nneth
          \\
          isomorphism
        \end{tabular}
      }
    }"{swap, yshift=-12pt}
  ]
  \ar[
    rrr,
    rounded corners,
    to path={
           ([yshift=+00pt]\tikztostart.south)  
        -- ([yshift=-12pt]\tikztostart.south)  
        -- node[yshift=-5pt]{
          \scalebox{.7}{
            \color{darkgreen}
            \bf
            Pontrjagin product
          }
        }
           ([yshift=-12pt]\tikztotarget.south)  
        -- ([yshift=-00pt]\tikztotarget.south)  
    }
  ]
  &  
  H_\bullet\big(
    \Omega \, \mathrm{Fields}
    \,\times\,
    \Omega \, \mathrm{Fields}
    ;\,
    \mathbb{C}
  \big)
  \ar[
    rr,
    "{
      \mathrm{conc}_\ast
    }",
    "{
      \scalebox{.6}{
        \color{darkgreen}
        \bf
        \def\arraystretch{.9}
        \begin{tabular}{c}
          push-forward along
          \\
          loop concatenation
        \end{tabular}
      }
    }"{swap}
  ]
  &\phantom{---}&
  H_\bullet\big(
    \Omega \, \mathrm{Fields}
    ;\,
    \mathbb{C}
  \big)
  \mathrlap{\,,}
  \end{tikzcd}
$$
and whose star-involution (antipode) is the push-forward along the operation ``$\mathrm{rev}$'' of reversal of loops followed by complex conjugation of the coeffients:

\vspace{-.7cm}
$$
  \begin{tikzcd}
    H_\bullet\big(
      \Omega\, 
      \mathrm{Fields}
      ;\,
      \mathbb{C}
    \big)
    \ar[
      rr,
      "{
        \mathrm{rev}_\ast
      }",
      "{
        \scalebox{.6}{
          \color{darkgreen}
          \bf
          \def\arraystretch{.9}
          \begin{tabular}{c}
            push-forward along
            \\
            loop reversal
          \end{tabular}
        }
      }"{swap}
    ]
  \ar[
    rrrr,
    rounded corners,
    to path={
           ([yshift=+00pt]\tikztostart.south)  
        -- ([yshift=-11pt]\tikztostart.south)  
        -- node[yshift=-5pt]{
          \scalebox{.7}{
            \color{darkgreen}
            \bf
            Pontrjagin star-involution
          }
        }
           ([yshift=-11pt]\tikztotarget.south)  
        -- ([yshift=-00pt]\tikztotarget.south)  
    }
  ]
    &\phantom{--}&
    H_\bullet\big(
      \Omega\, 
      \mathrm{Fields}
      ;\,
      \mathbb{C}
    \big)
    \ar[
      rr,
      "{ (-)^\ast }",
      "{
        \scalebox{.6}{
          \color{darkgreen}
          \bf
          \def\arraystretch{.9}
          \begin{tabular}{c}
            complex conjugation
          \end{tabular}
        }
      }"{swap}
    ]
    &\phantom{--}&
    H_\bullet\big(
      \Omega\, 
      \mathrm{Fields}
      ;\,
      \mathbb{C}
    \big)
  \end{tikzcd}
  \hspace{.5cm}
  \adjustbox{
    raise=-1.3cm
  }{
\begin{tikzpicture}[
    xscale=1.1,
    CoilColor/.store in=\coilcolor,CoilColor=black,
    Step/.store in=\Step,Step=0.1,
    Width/.store in=\Width,Width=0.4,
    Coil2/.style={
        decorate,
        decoration={
            markings,
            mark= between positions 0 and 1 step \Step 
            with {
                \begin{scope}[yscale=#1]
                    \pgfmathparse{int(\pgfdecoratedpathlength/28.45*100*\Step)}
                    \edef\Hight{\pgfmathresult}
                    \ifnum\pgfkeysvalueof{/pgf/decoration/mark info/sequence number}=1
                        \path (0,0)++(90: \Hight/200 and \Width) coordinate (b);
                    \fi
                    \ifnum\pgfkeysvalueof{/pgf/decoration/mark info/sequence number}>1
                        \coordinate (b) at (d);
                    \fi
                    \path (b) arc (90:-135: \Hight/200 and \Width) coordinate (a);
                    \path (b) arc (90:-45: \Hight/200 and \Width) coordinate (c);
                    \path (b)++(\Hight/100,0) coordinate (d);
                    \draw[fill,\coilcolor!70!black]
                        (c)
                            .. controls +(-0.175,0) and +(-0.275,0) .. (d)
                            .. controls +(-0.325,0) and +(-0.225,0) .. (c);
                    \draw[white,line width=2pt]
                        (b)
                            .. controls +(0.3,0) and +(0.2,0) .. (c);
                    \draw[fill,\coilcolor]
                        (b)
                            .. controls +(0.275,0) and +(0.175,0) .. (c)
                            .. controls +(0.225,0) and +(0.325,0) .. (b);
                \end{scope}
            }
        }
    }
]

 \draw[
   Coil2=2,
   CoilColor=black,
   Step=0.52,
  ] 
  (0,-.27) -- ++ (0,.75);

\draw[
  white,
  line width=1.2
]
  (-.1,-.4)
    ellipse
  (.804 and .2);
\draw
  (-.1,-.4)
    ellipse
  (.804 and .2);

 \draw[
   Coil2=2,
   CoilColor=black,
   Step=0.52,
  ] 
  (0,-1.11)  -- ++ (0,.75);

  \draw[-Latex]
    (-.9, -1.2)
    to node[yshift=5pt, xshift=4pt, sloped, pos=.7] {\scalebox{.6}{time}}
    (-.9,.8);

\draw[
  bend left=11,
  decorate,
  decoration={
    text along path,
    text align=center,
    text={|\tiny|
      periodic space
  }}
]
  (-.9,-.43) to (.7,-.445);
  
\end{tikzpicture}
}
$$

In algebraic quantum theory such a star-algebra may be understood as an algebra of quantum observables (cf. \cite[\S 2.2]{CSS23}) where the operator-ordering corresponds to time-ordering (\cite[p. 35]{Feynman42}\cite[p. 381]{Feynman48}, cf. \cite[pp. 33]{Nagaosa99}) and the star-involution reflects time-reversal. Therefore we may think of \eqref{QuantumObservablesOnM5} as being a topological-sector light-cone quantization, namely in a frame where the system continuously loops around the M-theory circle while evolving in time.

\smallskip

In the degree=0 that we will be focusing on here, all this simplifies: The degree=0 Pontrjagin algebra is just the complex {\it group algebra} \cite{Passman76} of the fundamental group of the field moduli:
\begin{equation}
  \label{ObservablesGiveGroupAlgebra}
  H_0(
    \Omega\, \mathrm{Fields}
    ;\,
    \mathbb{C}
  )
  \;\;
  \simeq
  \;\;
  \mathbb{C}[
    \pi_0 \, \Omega\, \mathrm{Fields}
  ]
  \;\;
  \simeq
  \;\;
  \mathbb{C}[
    \pi_1 \mathrm{Fields}
  ]
  \,.
\end{equation}
In the case at hand 
\eqref{QuantumObservablesOnM5}
is the group algebra of the third homotopy group of the 2-sphere
\begin{equation}
  \label{FundamentalGroupOfModuliSpace}
  \pi_0
  \Big(
    \Omega
    \,
    \mathrm{Maps}\big(
      \mathbb{R}^{2}_{\cpt}
      ,\,
      S^2
    \big)
  \Big)
  \;\simeq\;
  \pi_0
  \Big(
    \mathrm{Maps}\big(
      \mathbb{R}^1_{\cpt}
      \wedge
      \mathbb{R}^{2}_{\cpt}
    ;\,
    \mathbb{C}
  \Big)
  \;\simeq\;
  \pi_0
  \Big(
    \mathrm{Maps}\big(
      S^3
      ,\,
      S^2
    \big)
  \Big)
  \;\simeq\;
  \pi_3(S^2)
  \;\simeq\;
  \mathbb{Z}
  \,,
\end{equation}
which is generated by the framed link representing the Hopf fibration, which is
the unit-framed unknot (by \cite[(26)]{SS24-AbAnyons}).

\medskip

\noindent
{\bf Anyonic quantum states.}
Now one finds \cite[Thm. 3.18]{SS24-AbAnyons} that the framed link diagrams $L$ as indicated in \hyperlink{SolitonsBraiding}{\it Figure B} contribute through their total linking number $\# L$ (including self-linking/framing number) to \eqref{FundamentalGroupOfModuliSpace}:
\begin{equation}
  \label{TheTotalLinkingNumber}
  \begin{tikzcd}[row sep=-3pt, column sep=0pt]
    \Omega
    \,
    \mathrm{Maps}\big(
      \mathbb{R}^{2}_{\cpt}
      ,\,
      S^2
    \big)
  \ar[
    rr,
    ->>
  ]
  &&
  \pi_0
  \Big(
    \Omega
    \,
    \mathrm{Maps}\big(
      \mathbb{R}^{2}_{\cpt}
      ,\,
      S^2
    \big)
  \Big)
  \simeq
  \mathbb{Z}
  \\
  L &\longmapsto& 
  \quad 
\scalebox{.85}{$
  \# L
  \,:=\,
  \sum_i \mathrm{frm}(L_i)
  \,+\,
  \sum_{i,j} \mathrm{lnk}(L_i, L_j)
$  }
  \,,
  \end{tikzcd}
\end{equation}
where $(L_i)_{i = 1}^N$ denote the connected components of $L$.
But this is just the invariant seen by Chern-Simons Wilson-line observables 
(\cite[p. 363]{Witten89}, cf. review in \cite[(5.1)]{MPW19})
-- including their usual framing-regularization! 

More in detail: The {\it pure} quantum states $\vert k \rangle$ on an abelian star-algebra of quantum observables such as \eqref{FundamentalGroupOfModuliSpace} correspond (via their expectation values)  to algebra homomorphisms (cf. \cite[Ex. 13.3-4]{Zhu93}):
$$
  \begin{tikzcd}
    H_0\Big(
      \Omega
      \,
      \mathrm{Maps}\big(
        \mathbb{R}^2_{\cpt}
        ,\,
        S^2
      \big)
      ;\,
      \mathbb{C}
    \Big)
    \ar[
      rr,
      "{
        \langle k \vert 
        -
        \vert k \rangle
      }"
    ]
    &&
    \mathbb{C}
    \,,
  \end{tikzcd}
$$
and the group algebra on the left is generated by the class of the unit-framed unknot, such a state is fixed by the expectation value of this framed link, which by the reality and normalizability condition on states is some unimodular complex number $\exp(2\pi\mathrm{i}/k)$. With this,  the expectation value of a general framed-link observables is hence (\cite[Prop. 4.3]{SS24-AbAnyons})
$$
  \langle k \vert 
    L
  \vert k \rangle
  \;=\;
  \exp\big(
    \tfrac{2 \pi \mathrm{i}}{k}
    \,
    \# L
  \big)
  \;
  \underset{
    \mathclap{
      \adjustbox{
        raise=-3pt,
        scale=.7
      }{
        \eqref{TheTotalLinkingNumber}
      }
    }
  }{=}
  \;
  \exp\Big(
    \tfrac{2\pi\mathrm{i}}{k}
    \big(
      \sum_i \mathrm{frm}(L_i)
      \,+\,
      \sum_{i,j} \mathrm{lnk}(L_i, L_j)
    \big)
  \Big)
  \,.
$$
These expressions are exactly (\cite[Rem. 4.5]{SS24-AbAnyons}) the properly regularized expectation values of the Wilson loop $L$ in $\mathrm{U}(1)$-Chern-Simons theory, thus identifying our quantized solitonic field configurations, tracing out these links $L$, as being abelian anyons.

\section{Conclusion \& Outlook}

On the backdrop of arguments that M-theory -- and here specifically the worldvolume theory of M5-branes -- could provide the missing microscopic theory for strongly-coupled/correlated quantum systems -- such as topologically ordered quantum materials with anyonic quantum states --, we have turned attention to the first phenomenon to be taken care of when completing 11D supergravity by non-perturbative effects: namely the global solitonic completion of its usual local field content by flux quantization laws in (nonabelian) generalized cohomology. 

\smallskip 
By systematic discussion of the super-space Bianchi identities for the ``self-dual'' tensor field on M5 orbi-worldvolumes,
following \cite{GSS24-FluxOnM5},
and invoking recent mathematical results 
\cite{FSS23-Char}\cite{SS25-EquTwistorial}
on the relevant low-dimensional algebraic topology for identifying and analyzing its somewhat subtle flux quantization laws that are actually admissible (not just in the ``decoupling limit'' from the bulk SuGra), we identified one flux quantization law that is both technically admissible as well as plausible in that it satisfies a whole list of checks reproducing subtle expected topological effects in M-theory: this is the ``equivariant twistorial Cohomotopy'' theory developed in \cite{SS25-EquTwistorial}, in further refinement of the twistorial Cohomotopy of  \cite{FSS22-Twistorial} and the twisted Cohomotopy of \cite{FSS20-H}\cite{FSS21-Hopf}.

\smallskip 
The key observation we made then is that 
on trivially Seifert-fibered M5 orbi-worldcolumes this flux quantization in twistorial equivariant Cohomotopy has the remarkable property of  reducing to plain 2-Cohomotopy on the 1+2-dimensional orbi-singularity (times the M-theory circle). This allowed to next invoke recent results of \cite{SS24-AbAnyons} to rigorously deduce abelian anyonic quantum states of the resulting solitonic field configurations in 1+2D. 

\smallskip

We suggest that this result may point the way to break the theoretical impasse both in strongly-coupled quantum systems (where true microscopic understanding tends to be missing) and in the development of M-theory (where the web of conjectures has been lacking anchoring in a tangible foundation) and in their combination by holographic duality (where the unrealistic limit of a large number $N$ of ``coincident'' branes has been serving as a crutch): 

\smallskip
The result indicates that the systematic flux-quantization of M-branes probing 11D SuGra backgrounds -- largely ignored in previous literature -- is a substantial first step in the notoriously necessary but previously elusive ``completion'' of 11D SuGra towards M-theory, and hence potentially towards a first-principles theory of strongly-coupled quantum systems.

\end{document}